# Bloch oscillations of Fibonacci anyons

Xiaoqi Zhou[*], Weixuan Zhang[*, +], Hao Yuan, and Xiangdong Zhang[$]

*Key Laboratory of advanced optoelectronic quantum architecture and measurements of Ministry of Education, Beijing Key Laboratory of Nanophotonics & Ultrafine Optoelectronic Systems, School of Physics, Beijing Institute of Technology, 100081, Beijing, China*

*These authors contributed equally to this work.

[$+]Author to whom any correspondence should be addressed: zhangxd@bit.edu.cn; zhangwx@bit.edu.cn

**ABSTRACT**

**Non-Abelian anyons, which correspond to collective excitations possessing multiple fusion channels and noncommuting braiding statistics, serve as the fundamental constituents for topological quantum computation. Here, we reveal the exotic Bloch oscillations (BOs) induced by non-Abelian fusion of Fibonacci anyons. It is shown that the interplay between fusion-dependent internal energy levels and external forces can induce BOs and Bloch-Zener oscillations (BZOs) of coupled fusion degrees with varying periods. In this case, the golden ratio of the fusion matrix can be determined by the period of BOs or BZOs in conjunction with external forces, giving rise to an effective way to unravel non-Abelian fusion. Furthermore, we experimentally simulate non-Abelian fusion BOs by mapping Schrödinger equation of two Fibonacci anyons onto dynamical equation of electric circuits. Through the measurement of impedance spectra and voltage evolution, both fusion-dependent BZOs and BOs are simulated. Our findings establish a connection between BOs and non-Abelian fusion, providing a versatile platform for simulating numerous intriguing phenomena associated with non-Abelian physics.**

# I. INTRODUCTION

Quantum computing has garnered significant attention in recent decades due to its potential for exponential acceleration over classical computers in solving specific problems. The principal challenges in the development of quantum computing lie in addressing noise and decoherence. Topological quantum computation presents a promising approach to solving these obstacles by utilizing non-Abelian anyons for encoding and manipulating quantum qubits in a nonlocal manner, where quantum gates are implemented by braiding of non-Abelian anyons and the measurement of quantum qubits is accomplished through fusion of anyons [1-5]. In contrast to fermions and bosons, the exchange of non-Abelian anyons is governed by unitary transformations. In addition, there are multiple possible outcomes upon fusion (merging) of two non-Abelian anyons, being different from other elementary particles that have only one fusion outcome. These two fundamental properties of non-Abelian anyons form the basis for topological quantum computation. Given their significant applications, extensive research has been conducted to identify physical systems capable of hosting non-Abelian anyons.

Two promising platforms for non-Abelian anyons are the Fibonacci and Ising anyon models, both exhibiting non-Abelian braiding statistics and fusion roles [1-25]. Specifically, Ising anyons have been predicted to exist as Majorana zero modes in topological superconductors [6-9], low-energy excitations of fractional quantum Hall states [12,13], Kitaev materials [14-16] and dense quark matter [17]. In addition, the vortex-bounded Dirac fermionic mode is also proposed to be among the Ising anyons [18-22]. However, Ising anyons alone are insufficient for universal quantum computations, therefore, incorporating non-topological processes becomes necessary [2]. In contrast, Fibonacci anyons offer a promising platform for universal topological quantum computation where all quantum gates are implemented through braiding manipulation in a topologically protected manner [23], and several candidates have been proposed to host Fibonacci anyons, including the fractional quantum Hall state with $\nu = 12/5$ [24], interacting Majorana fermions [25], and so on [26-30]. In addition, some recent investigations have shown that anyonic statistics could trigger the appearance of many interesting effects [31-34]. In particular, it has been pointed out that quantum statistics could dramatically affect the Bloch oscillation (BO) of two Abelian anyons, where the oscillation frequency of two pseudofermions with a statistical angle of $\pi$ becomes half of that for two bosons [31, 32]. Motivated by the novel properties of anyonic BOs, it is intriguing to inquire whether non-Abelian statistics can give rise to more exotic BOs.

In this work, we firstly investigate the non-Abelian fusion BOs and Bloch-Zener oscillations (BZOs)

of Fibonacci anyons in a one-dimensional lattice subject to external forces. It is shown that by adjusting the external forcing, it is possible to achieve either two or a single Wannier-Stack ladder of two Fibonacci anyons with a fusion-dependent internal energy level. When a single Wannier-Stack ladder accompanied with BOs emerges, the corresponding values of external force can determine the golden ratio for the fusion matrix of Fibonacci anyons. As for the case of three Fibonacci anyons, there are a pair of internal energy levels, making BZOs appear under a suitable value of external force. Furthermore, we design and fabricate RC circuits to experimentally simulate BOs and BZOs of Fibonacci anyons. Our work establishes a connection between BOs dynamics and non-Abelian fusion, suggesting a useful way to characterize non-Abelian fusion.

## II. THE THEORY OF CHARACTERIZING NON-ABELIAN FUSION OF FIBONACCI ANYONS BY BLOCH OSCILLATIONS AND BLOCH-ZENER OSCILLATIONS.

Previous investigations have shown that Fibonacci anyons can be described by the $SU(2)_3$ Chern-Simons theories [35, 36] according to the fusion rule of $\tau \otimes \tau = \tau \oplus I$ with $\tau$ and $I$ representing a Fibonacci anyon and a vacuum state, respectively. In recent years, the lattice models of Fibonacci anyons with finite filling factors have been wildly explored, giving rise to various exotic many-body phenomena related to non-Abelian fusion [37-40]. Here, following the $SU(2)_3$ theory with the above fusion role, we consider a pair of Fibonacci anyons moving along a one-dimensional (1D) chain under the influence of an external force $F$. Specifically, the system consists of two Fibonacci anyons and $L$-2 vacuum states, where $L$ represents the length of the lattice, as depicted in Fig. 1(a). $J_t$ is the single-anyon hopping rate, and $J$ is the interaction strength of Fibonacci anyons. The label $x_l$ corresponds to a non-Abelian anyon resulting from combining $x_{l-1}$ with its preceding non-Abelian anyon at the ($l$-1)th lattice site as $A_{l-1} \otimes x_{l-1} = x_l$ where $A_{l-1}$ can be either the vacuum particle $I$ or the non-Abelian anyon $\tau$. In this case, different two-anyon states within the Hilbert space should not only be characterized by positions of Fibonacci anyons but also by labels $x_l$. There are five possible values of $x_1, x_2, ..., x_{L+1}$ when positions of two Fibonacci anyons are fixed (see Appendix A for details). Therefore, each two-anyon configuration in our model exhibits five internal degrees of freedom arising from the fusion of two Fibonacci anyons, being called as fusion degrees. The probability amplitude of two-anyon state with one Fibonacci anyon at the $m$th lattice site and the other at the $n$th lattice site, along with the $\alpha$th fusion degree ($\alpha$=1, 2, 3, 4, 5), can be denoted as $C_{m,n}^\alpha$.

In addition, Fibonacci anyons in our model possess the effective interaction ($J$), which can be quantized by the $\mathcal{F}$-symbol transformation, serving as a basis transformation between different representations of fusion trees [37-40]. By applying the $\mathcal{F}$-symbol transformation, we can transform the fusion chain basis depicted in Fig. 1(a) to a basis that exhibits the direct fusion channel of two Fibonacci anyons, as indicated in Fig. 1(b) with the label $\widetilde{x_L}$. Thus, analogous to the Heisenberg interaction of electrons, the interaction among Fibonacci anyons can be obtained by projecting the fusion outcome onto a vacuum state $I$. Thus, the fusion matrix is denoted by

$$H_J = -J \begin{pmatrix} 1 & 0 & 0 & 0 & 0 \\ 0 & 0 & 0 & 0 & 0 \\ 0 & 0 & 0 & 0 & 0 \\ 0 & 0 & 0 & \phi^{-2} & \phi^{-3/2} \\ 0 & 0 & 0 & \phi^{-3/2} & \phi^{-1} \end{pmatrix}, \quad (1)$$

where $\phi = (\sqrt{5}+1)/2$ is the golden ratio of the fusion matrix. Details for the derivation of the fusion matrix $H_J$ is provided in Appendix B. It is noteworthy that the external force $F$ can act on each non-Abelian fusion degree. In this scenario, the impact of external force can be reinterpreted as the gradient potential $F(m+n)I_{5\times 5}$ with two Fibonacci anyons at the $m$th and $n$th sites.

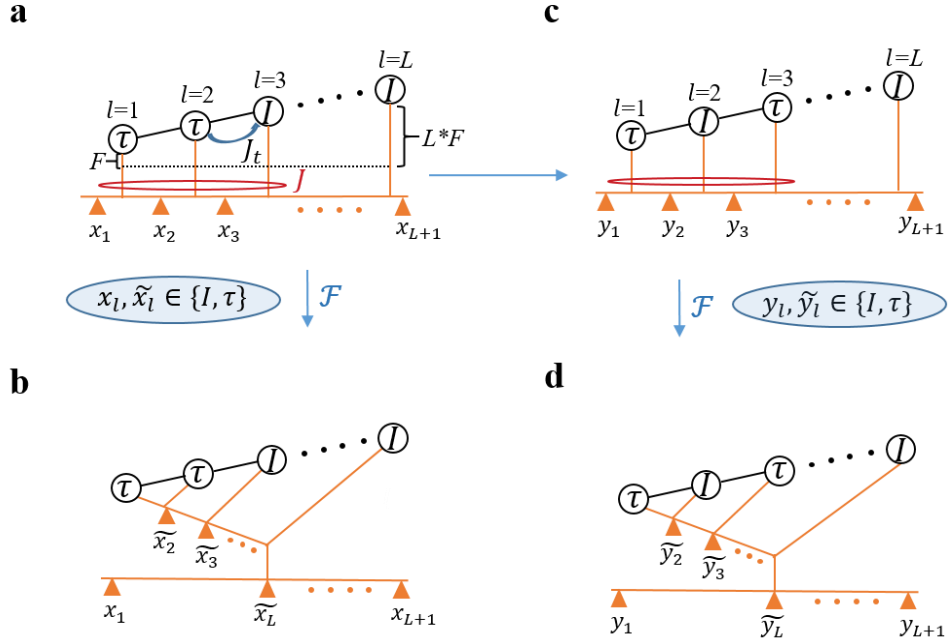

FIG. 1. The schematic diagram for the lattice model of two Fibonacci anyons under the external force $F$. (a) The fusion tree with a Fibonacci anyon at the first site and the other Fibonacci anyon at the second site. (b) By applying the $\mathcal{F}$-symbol transformation, the fusion chain basis is transformed into a basis that displays the direct fusion channel of two Fibonacci anyons located at the first and second positions. (c) The fusion tree corresponding to a Fibonacci anyon at the first site and the other Fibonacci anyon at the third site. (d) By applying the $\mathcal{F}$-symbol transformation, the fusion chain basis is transformed into a basis that displays the direct fusion channel of two Fibonacci anyons located at the first and third positions. The label $x_l$ ($y_l$) corresponds to the outcome resulting

from merging $x_{l-1}$ ($y_{l-1}$) with its preceding non-Abelian anyon at ($l$-1)th site $A_{l-1}$, being expressed as $A_{l-1} \otimes x_{l-1}(y_{l-1}) = x_l$ ($y_l$) where $A_{l-1}$ can be either the vacuum particle $I$ or the non-Abelian anyon $\tau$.

Except for the fusion interaction of Fibonacci anyons, each anyon can also transport from one site to an adjacent vacant site. In particular, when a Fibonacci anyon is hopping from the second site to the third site, the corresponding fusion tree of the final state is also changed, as illustrated in Fig. 1(c). Similarly, we can apply the $\mathcal{F}$-symbol to transform the fusion chain basis represented in Fig. 1(c) to the basis that displays the direct fusion channel of two Fibonacci anyons, as indicated in Fig. 1(d). In this case, the coupling matrix between the initial and final two-anyon states is described by $\boldsymbol{H}_t = -J_t \boldsymbol{I}_{5 \times 5}$. Thus, the time-dependent Schrödinger equation of two Fibonacci anyons can be expressed as ($\hbar = 1$):

$$i\frac{d}{dt}\boldsymbol{\psi}_{m,n} = \boldsymbol{H}_t[\boldsymbol{\psi}_{m\pm 1,n} + \boldsymbol{\psi}_{m,n\pm 1}] + [\boldsymbol{H}_J + F(m+n)\boldsymbol{I}_{5\times 5}]\boldsymbol{\psi}_{m,n} \quad (2)$$

with $\boldsymbol{\psi}_{m,n} = [C_{m,n}^1, C_{m,n}^2, C_{m,n}^3, C_{m,n}^4, C_{m,n}^5]^T$. Here, $C_{m,n}^\alpha$ corresponds to the probability amplitude with one Fibonacci anyon at the mth site and the other at the nth site, along with the αth fusion degree (α=1, 2, 3, 4, 5). $\boldsymbol{\psi}_{m,n}$ is the state vector composed of $C_{m,n}^\alpha$ for five fusion degrees. $\boldsymbol{H}_t = -J_t \boldsymbol{I}_{5\times 5}$ is the coupling matrix between a pair of two-anyon states with $J_t$ being the single-anyon hopping rate. Specifically, the first two terms on the right-hand side of Eq. (2) correspond to scenarios where one of two Fibonacci anyons transitions from the mth site to either the (*m*+1)th or (*m*-1)th site, or from the nth site to either the (*n*+1)th or (*n*-1)th site. $\boldsymbol{H}_J$ is the fusion matrix. The impact of the external force $F$ can be reinterpreted as the gradient potential $F(m+n)\boldsymbol{I}_{5\times 5}$.

Next, we study the eigenspectra and dynamics of two Fibonacci anyons based on Eq. (2). It is noted that fusion degrees of $C_{m,n}^1$, $C_{m,n}^2$ and $C_{m,n}^3$ are all decoupled from other fusion degrees. In this case, two Fibonacci anyons with these three fusion degrees behave as two non-interacting particles hopping on a 1D chain under an external force $F$, with the effective onsite potential of each particle being –*J*, 0 and 0 for the case of $C_{m,n}^1$, $C_{m,n}^2$ and $C_{m,n}^3$, respectively. In this case, the two-anyon eigenspectra can exhibit the Wannier-Stark ladder $\varepsilon_{mn}^1 = -J + (m+n)F$ related to $C_{m,n}^1$ and $\varepsilon_{mn}^2 = \varepsilon_{mn}^3 = (m+n)F$ for $C_{m,n}^2$ and $C_{m,n}^3$. The difference between two adjacent eigen-energies equals to $\delta\varepsilon_{1,2,3} = F$ (see Appendix C for the numerical results of $\varepsilon_{mn}^1$ as a function of *F*).

In contrast to the first, second, and third fusion degrees, it is noted that the fourth and fifth fusion degrees are intricately coupled with each other, involving the golden ratio $\phi$ associated with Fibonacci anyons. In such a case, both the external force and non-Abelian fusion can influence the energy levels

$\varepsilon_{m,n}^{4,5}$ related to two non-separable fusion degrees $C_{m,n}^4$ and $C_{m,n}^5$ for two Fibonacci anyons. To clarify the property of the energy level for $\varepsilon_{m,n}^{4,5}$, we first consider the case with $F=0$. The interval energy level of fourth and fifth fusion degrees can be obtained by solving eigenvalues of $H_J^{4,5} = -J\begin{pmatrix} \phi^{-2} & \phi^{-3/2} \\ \phi^{-3/2} & \phi^{-1} \end{pmatrix}$, being $\varepsilon_a^{4,5} = 0$ and $\varepsilon_b^{4,5} = 1$. Thus, the fusion-dependent internal energy gap of $\varepsilon_{m,n}^{4,5}$ equals to $\delta\varepsilon_f^{4,5} = 1$. Hence, two distinct Wannier-Stack ladders with energy gaps being $\delta\varepsilon_{4,5}^1 = \varepsilon_b^{4,5} - \varepsilon_a^{4,5}$ and $\delta\varepsilon_{4,5}^2 = F - (\varepsilon_b^{4,5} - \varepsilon_a^{4,5})$ are obtained with $F \geqslant 1$. We anticipate the occurrence of BZOs with two non-coincident Wannier-Stack ladders [41]. In other words, the presence of two-anyon BZOs would serve as a confirmation for the existence of the non-Abelian fusion. The period of BZOs, denoted as $T_{BZO}$, equals to the least common multiple of $T_1 = 2\pi/\delta\varepsilon_{4,5}^1$ and $T_2 = 2\pi/\delta\varepsilon_{4,5}^2$. Interestingly, it is noted that two Wannier-Stack ladders can also form a superposition pattern with $\delta\varepsilon_{4,5}^1 = \delta\varepsilon_{4,5}^2$ at $F=0.5$, 1, and 2 (See Appendix C), triggering the appearance of BOs [42-45] for two coupled fusion degrees. Based on the values of external forces that can trigger BOs, we can determine the golden ratio $\phi$ for the fusion matrix of Fibonacci anyons, making BOs act as an effective way to resolve the non-Abelian fusion.

To demonstrate the emergence of BZOs and BOs resulting from fusion energy levels of $\varepsilon_{mn}^{4,5}$ at different external forces, we calculate the temporal evolution of two-anyon states $C_{m,n}^4$ and $C_{m,n}^5$ with $L=20$ and $J_t=J=-1$. At first, we focus on the BZO in a system with two staggered Wannier-Stack ladders under $F=1.5$. We initialize the two-anyon state as $C_{6,13}^5 = 1$ and calculate the evolution of the system. Fig. 2(a) presents the calculated wave dynamics of $C_{6,13}^5$. It is clearly shown that the periodic oscillation appears with $T_{BZO} = 12.56$, being consistent to the above theoretical prediction. A distinctive feature in the Fibonacci anyon system is the emergence of coupled fusion degrees, making these two fusion degrees exhibit coupled BZOs. To illustrate their oscillatory behaviors, we calculate the time-dependent particle occupation density $P_i^\alpha = \sum_{l=1}^L |C_{i,l}^\alpha|^2$ ($i \neq l, \alpha = 4,5$) at different times in Fig. 2(b). Black and red lines correspond to particle occupation densities associated with the fourth and fifth fusion degrees, respectively. At $t=0.3$ (the red triangle), the occupation density is observed to be highest for the fifth fusion degree at the sixth and thirteenth lattice sites, with a low particle occupation density of the fourth fusion degree. As time increases to $t=4.43$ (the blue triangle), the occupation density reaches its peak for the fourth fusion degree at the sixth and thirteenth lattice sites, while the occupation density of the fifth fusion degree is decreased. When $t=6.8$ (the yellow triangle), the maximum oscillation range is achieved, with both fusion degrees exhibiting their highest occupation densities at the fifth, seventh, twelfth and fourteenth lattice points. Additionally, a significant portion of the occupation density has transformed to the fifth fusion degree. Upon reaching $T_{BZ} = 12.56$ (the green triangle), which marks one complete period, it is evident that both fusion degrees return to the initial distribution. This periodic modulation in the probability distribution not only supports our prior theoretical conjectures concerning BZO of Fibonacci anyons, but also substantiates the existence of fusion energy levels within them.

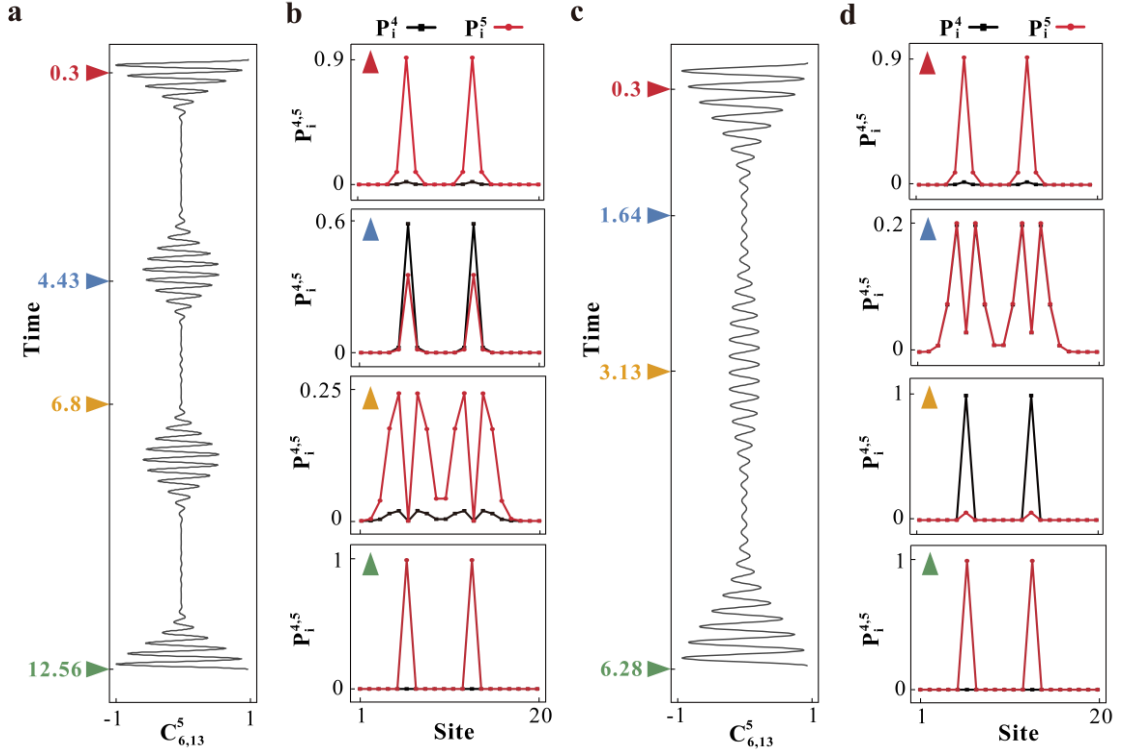

**FIG. 2. Numerical results of the BZO and BO for two Fibonacci anyons.** (**a**) The calculated wave dynamics of the probability amplitude $C_{6,13}^5$ with the external force being $F=1.5$. (**b**) The evolution of the particle occupation density $P_i^\alpha = \sum_{l=1}^L |C_{i,l}^\alpha|^2$ ($i \neq l, \alpha = 4, 5$) at different times with $F=1.5$. The black and red lines represent the fourth and fifth fusion degrees, respectively. (**c**) The calculated wave dynamics of probability amplitude $C_{6,13}^5$ with the external force being $F=2$. (**d**) The evolution of the particle occupation density at different times with $F=2$.

Next, we direct our attention to the behavior of two-anyon BOs with $F=2$. We initialize the system with $C_{6,13}^5 = 1$. Fig. 2(c) illustrates the temporal evolution of the probability amplitude $C_{6,13}^5$. The calculated period of BOs is found to be 6.28, in agreement with the predicted value. Furthermore, we evaluate the evolution of particle occupation densities to observe the mutual transformation between two coupled fusion degrees during BOs, as depicted in Fig. 2(d). At $t=0.3$, the occupation density is highest for the fifth fusion degree at the sixth and thirteenth lattice site, while the occupation density for the fourth fusion degree starts to increase. A continuous transformation occurs between these two fusion degrees, and at $t=1.64$, both degrees of freedom exhibit nearly equal occupation densities, reaching maximum values at the fifth, seventh, twelfth and fourteenth lattice points. At this moment, BOs of two fusion degrees reach their maximum range. As time increases to 3.134, the transformation between these mutually coupled degrees reaches its peak level. The occupation density is highest for the fourth fusion degree at lattice points six and thirteen, while it decreases for the fifth fusion degree of freedom at those same lattice points. When time reaches to $T_{BO} = 6.28$, we observe that the initial state oscillates back to sixth and thirteenth lattice points with an occupation density being zero for the fourth fusion degree, confirming our previous theoretical prediction. Above results clearly show that the reciprocal transformations among coupled fusion degrees appear during BZOs and BOs of two Fibonacci anyons.

It is noticed that the fusion degrees in the 1D chain are related to the number of Fibonacci anyons. Thus, we further consider the lattice model with three Fibonacci anyons. The fusion matrix of three Fibonacci anyons can be derived as (see Appendix D for details):

$$H_{3,J} = -J \begin{pmatrix} 1 & 0 & 0 & 0 & 0 & 0 & 0 & 0 \\ 0 & 0 & 0 & 0 & 0 & 0 & 0 & 0 \\ 0 & 0 & 0 & 0 & 0 & 0 & 0 & 0 \\ 0 & 0 & 0 & 0 & 0 & 0 & 0 & 0 \\ 0 & 0 & 0 & 0 & 0 & 0 & 0 & 0 \\ 0 & 0 & 0 & 0 & 0 & \phi^{-2} & \phi^{-2} & -\phi^{5/2} \\ 0 & 0 & 0 & 0 & 0 & \phi^{-2} & \phi^{-2} & -\phi^{-5/2} \\ 0 & 0 & 0 & 0 & 0 & -\phi^{-5/2} & -\phi^{-5/2} & -\phi^{-3} \end{pmatrix}, \tag{3}$$

The coupling matrix of the three-anyon model is $H_{3,t} = -J_t I_{8\times 8}$. Here, $C_{m,n,q}^{\alpha}$ is used to represent the probability amplitude of the $\alpha$th fusion degree when the first anyon is at position $m$, the second anyon is at the position $n$, and the third anyon is at the position $q$. The time-dependent Schrödinger equation of three-anyon state $C_{m,n,q}^{\alpha}$ can be found in Appendix E.

For the three Fibonacci anyon system, we find that there are eight fusion degrees, where the first five fusion degrees are decoupled from others and last three degrees are coupled together. In this case, we focus on analyzing the impact of three coupled fusion degrees on the BZO. The eigenvalues of

$H_{3,J}^{6,7,8} = -J \begin{pmatrix} \phi^{-2} & \phi^{-2} & -\phi^{5/2} \\ \phi^{-2} & \phi^{-2} & -\phi^{-5/2} \\ -\phi^{-5/2} & -\phi^{-5/2} & -\phi^{-3} \end{pmatrix}$ are equal to $\varepsilon_1 = -0.8937$, $\varepsilon_2 = 0$ and $\varepsilon_3 = 1.4216$, making the internal fusion energy levels become $\delta\varepsilon_f^1 = 0.8937$ and $\delta\varepsilon_f^2 = 1.4216$. In this case, two staggered Wannier-Stack ladders with energy gaps being $\Delta E_1 = F - \delta\varepsilon_f^2 = \delta\varepsilon_f^1$ and $\Delta E_2 = F - \delta\varepsilon_f^1 = \delta\varepsilon_f^2$ can appear with $F$=2.3153 (see Appendix E for details). Thus, it is anticipated that the BZOs can appear in the three-Fibonacci anyon systems with the oscillation period being $T_{BZO} = r\frac{2\pi}{\Delta E_1} = s\frac{2\pi}{\Delta E_2}$ (where $r$ and $s$ are a pair of prime numbers).

We calculate the evolution of three coupled fusion degrees with the initial state being $C^6_{5,10,15} = 1$. Fig. 3(a) illustrates the temporal evolution of $C^6_{5,10,15}$. Other parameters are $L=20$ and $J_t=J=-1$. It is shown that the oscillation period is equal to 35.26, which is consistent with the theoretical prediction. Within one period, the waveform is symmetric about $t = 17.63$. We compute the particle occupation densities of three fusion degrees at different times, as shown in Fig. 3(b). Black, red and blue lines correspond to results of the sixth, seventh and eighth fusion degrees, respectively. At $t = 0.3$, it can be observed that the sixth fusion degree has the highest particle occupation density, while the particle occupation densities of the seventh and eighth fusion degrees are nearly zero. As time increasing to 6.73, the occupation density of the sixth fusion degree has completely transformed into the occupation density of the seventh and eighth fusion degrees. At $t = 10.86$, the particle occupation density of the sixth fusion degree increases and the seventh fusion degree has a lower occupation density, while the particle density for the eighth fusion degree becomes zero. At $t=17.63$, the sixth and seventh fusion degrees have the highest particle occupation density, but the eighth degree of freedom has a lower particle occupation density. The mutual transformation of internal coupled fusion degrees during BZO provides a strong evidence for the existence of internal energy levels of the three Fibonacci anyon systems. It is worth noting that, as for other values of $F$, the system possesses three different internal energy levels, and the overall oscillation period should be the least common multiple of three periods related to these three energy levels (see Appendix F for details). In this scenario, an increase in the number of Fibonacci anyons leads to an increment of distinct internal energy levels, thereby significantly extending the period of fusion dynamics and disrupting the emergence of BOs and BZOs in the thermodynamic limit.

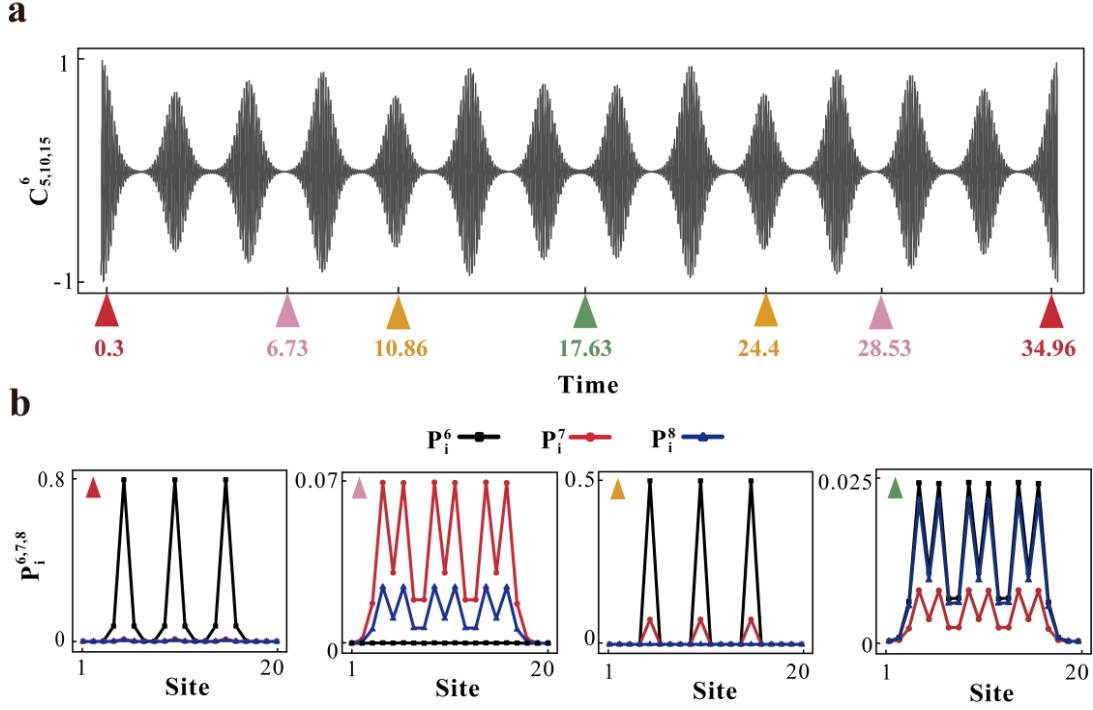

**FIG. 3. Numerical results of the BZO for three Fibonacci anyons.** (**a**) The temporal evolution of the probability amplitude $C^6_{5,10,15}$ with the external force being $F$=2.3153. (**b**) The evolution of the particle occupation densities of three fusion degrees at different times. Black, red and blue lines correspond to results of the sixth, seventh and eighth fusion degrees of three Fibonacci anyons, respectively.

## III. EXPERIMENTAL SIMULATION OF NON-ABELIAN FUSION OF FIBONACCI ANYONS DYNAMICS BY ELECTRIC CIRCUITS.

Motivated by previous experimental breakthroughs in simulating few-body quantum systems by electric circuits [32,46,47], optical waveguides [48] and fiber networks [49], in this part, we experimentally simulate the non-Abelian fusion BOs and BZOs of two Fibonacci anyons by mapping Schrödinger equation of two Fibonacci anyons (Eq. (2)) onto the dynamical equation of our designed RC circuits. Fig. 4(a) displays the schematic diagram of the RC circuit simulator. The voltages at pink and blue (gray and white) circuit nodes are labeled by $V^4_{m,n}$ and $V^5_{m,n}$ ($V'^4_{m,n}$ and $V'^5_{m,n}$), which can be mapped to the probability amplitudes $c^4_{m,n}$ and $c^5_{m,n}$ ($c'^4_{m,n}$ and $c'^5_{m,n}$) of a 1D lattice chain with two Fibonacci anyons. For example, the voltages $V^4_{1,2}$ and $V^5_{1,2}$ correspond to the probability amplitudes of the fourth and fifth fusion degrees with two Fibonacci anyons locating at the first and second lattice sites, as illustrated in the inset. The effective coupling between different two-anyon states can be realized by designing the node connection in the circuit. Fig. 4(b) plots the connection pattern from the circuit node $V^5_{3,5}$ to its nearby nodes of $V'^5_{2,5}$, $V'^5_{3,4}$, $V'^4_{3,5}$, $V'^5_{3,5}$, $V'^5_{3,6}$ and $V'^5_{4,5}$ through negative impedance converters

(INICs) valued of $\pm R_t$, $\pm R_t$, $\pm R_{J_2}$, $\pm R_{J_3} + R_F(m+n)$, $\pm R_t$ and $\pm R_t$. In addition, the circuit node $V_{3,5}^4$ is connected to $V_{2,5}^{\prime 4}$, $V_{3,4}^{\prime 4}$, $V_{3,5}^{\prime 5}$, $V_{3,5}^{\prime 4}$, $V_{3,6}^{\prime 4}$ and $V_{4,5}^{\prime 4}$ through $\pm R_t$, $\pm R_t$, $\pm R_{J_2}$, $\pm R_{J_1} + R_F(m+n)$, $\pm R_t$ and $\pm R_t$. Each node is also connected to the ground by a capacitor $C$. In this case, the eigen-equation of the voltage evolution can be derived as (see Appendix G for details)

$$\frac{d}{dt}V_{m,n} = \frac{1}{CR_t}\{H'_t[V'_{m\pm1,n} + V'_{m,n\pm1}] + \left[H'_J + \frac{1}{R_F(m+n)}I_{2\times 2}\right]V'_{m,n}\}$$

$$\frac{d}{dt}V'_{m,n} = -\frac{1}{CR_t}\{H'_t[V_{m\pm1,n} + V_{m,n\pm1}] + \left[H'_J + \frac{1}{R_F(m+n)}I_{2\times 2}\right]V_{m,n}\} \quad (4)$$

with $V_{m,n} = [V_{m,n}^4, V_{m,n}^5]^T$, $V'_{m,n} = [V_{m,n}^{\prime 4}, V_{m,n}^{\prime 5}]^T$, $H'_t = I_{2\times 2}$ and $H'_J = \begin{pmatrix} \frac{1}{R_{J_1}} & \frac{1}{R_{J_2}} \\ \frac{1}{R_{J_2}} & \frac{1}{R_{J_3}} \end{pmatrix}$. We can write the circuit eigen-equation of Eq. (4) in the matrix form of $i\partial_t|V(t)\rangle = \Pi|V(t)\rangle$ with $|V(t)\rangle = [V_{1,2}^4(t), V_{1,2}^5(t), \dots, V_{L-1,L}^4(t), V_{L-1,L}^5(t), V_{1,2}^{\prime 4}(t), V_{1,2}^{\prime 5}(t), \dots, V_{L-1,L}^{\prime 4}(t), V_{L-1,L}^{\prime 5}(t)]^T$ and $\Pi = i\begin{pmatrix} 0 & H \\ -H & 0 \end{pmatrix}$.

Here, if circuit elements satisfy $F = \frac{1}{CR_tR_F(m+n)}$, $\phi^{-2} = \frac{1}{CR_tR_{J_1}}$, $\phi^{-3/2} = \frac{1}{CR_tR_{J_2}}$, and $\phi^{-1} = \frac{1}{CR_tR_{J_3}}$, $H$ can be mapped to the Hamiltonian matrix of the fourth and fifth fusion degrees of two Fibonacci anyons. Thus, the eigen-energy $(\varepsilon_{mn}^{4,5})$ of two Fibonacci anyons is directly mapped to the eigen-frequency $(f)$ of the circuit as $\varepsilon_{mn}^{4,5} = 2\pi f R_t C$ (see Appendix G), making the designed RC circuit can be used to simulate the 1D lattice model with two Fibonacci anyons. Fig. 4(c) provides the photographic image of the fabricated circuit ($L$=7) around the node $V_{3,5}^5$ with $C = 1\mu F$, $R_t = 1000\Omega$, $R_{J_1} = 2618\Omega$, $R_{J_2} = 2058\Omega$, $R_{J_3} = 1618\Omega$ and $R_F = 1500\Omega$ ($F$=1.5). Details on the sample fabrication are provided in Appendix H.

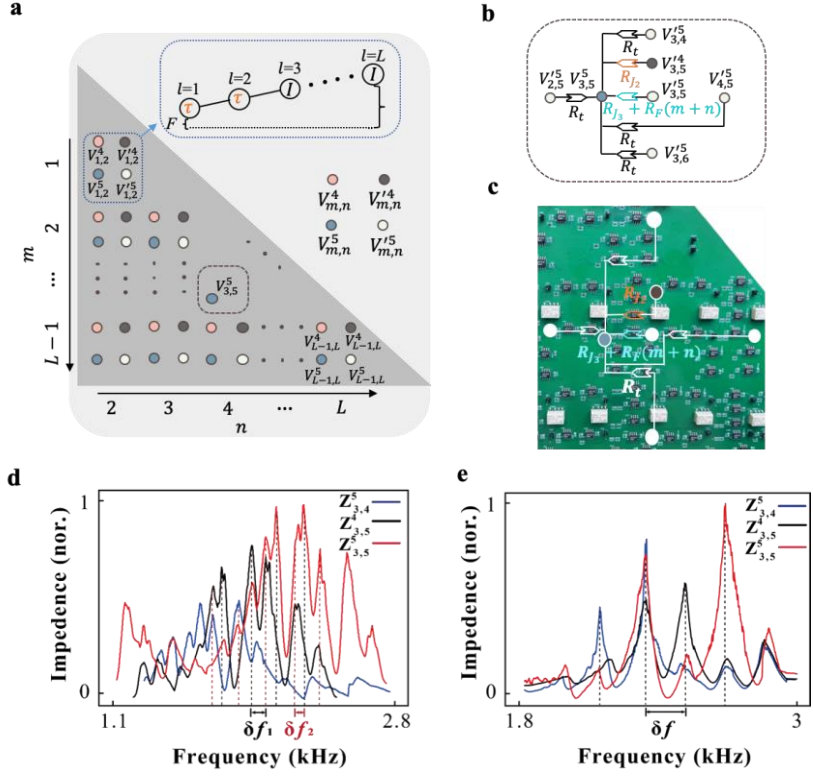

**FIG. 4. Experimental simulation of Wannier Stark ladders for two Fibonacci anyons by impedance spectra.** (**a**) The schematic diagram of the designed RC circuit simulator. The voltages at pink and blue (gray and white) circuit nodes are labeled by $V^4_{m,n}$ and $V^5_{m,n}$ ($V'^4_{m,n}$ and $V'^5_{m,n}$). The inset shows the correspondence between the voltages $V^4_{1,2}$ and $V^5_{1,2}$ and probability amplitudes of the fourth and fifth fusion degrees with two Fibonacci anyons locating at the first and second lattice sites. (**b**) The detailed connection pattern of the circuit node $V^5_{3,5}$ to nearby nodes through negative impedance converters. (**c**) The photographic image of the fabricated circuit (L=7) around the node $V^5_{3,5}$, with $C = 1\mu F$, $R_t = 1000\Omega$ (the white symbols), $R_{J_1} = 2618\Omega$, $R_{J_2} = 2058\Omega$ (the orange symbol), $R_{J_3} = 1618\Omega$ (the blue symbol), and $R_F = 1500\Omega$ (the blue symbol). Measured impedance spectra of $Z^5_{3,4}$, $Z^4_{3,5}$ and $Z^5_{3,5}$ for the circuit simulator with F=1.5 for (**d**), and F=2 for (**e**).

It is noted that the impedance response of a circuit node is related to the local density of states of the mapped quantum lattice model. Thus, to illustrate the eigen-spectrum of the circuit simulator, we measure impedance spectra of three circuit nodes ($V^5_{3,4}$, $V^4_{3,5}$ and $V^5_{3,5}$), as illustrated by blue, black and red lines in Fig. 4(d). It is shown that there are multiple impedance peaks of each circuit node, where each peak can be mapped to an eigenenergy of the two-anyon eigenspectrum. Interestingly, we can find that there are two intervals (marked by dash lines) between adjacent peaks, equaling to $\delta f_1$=0.15 kHz and $\delta f_2$=0.08 kHz, respectively. This is consistent with the existence of two Wannier Stark ladders in the mapped lattice model with $\delta\varepsilon_1$=1 and $\delta\varepsilon_2$=0.5. Compared to simulation results (see Appendix I), the large width of measured impedance peak is resulting from the loss effect of the circuit sample. Moreover, the frequency deviation of impedance peaks is attributed to disorder effect. In addition, we also fabricate the other circuit simulator with $R_F = 2000\Omega$, corresponding to *F*=2. Fig. 4(e) presents the measured impedance spectra of $Z_{3,4}^5$, $Z_{3,5}^4$ and $Z_{3,5}^5$. It can be observed that the interval between two adjacent impedance peaks are equally spaced with $\delta f$=0.159 kHz. This indicates the existence of a single Wannier Stark ladder in the mapped lattice model, being consistent with theoretical analysis.

To further simulate the fusion-dependent BZO and BO in our circuit simulators, we conduct time-domain measurements of voltage dynamics. Firstly, we measure the voltage evolution in the fabricated circuit with *F*=1.5. The initial voltage is set as $V_{3,5}^5(t=0) = 1V$. Fig. 5(a) displays the measured voltage signal of $V_{3,5}^5(t)$. It can be observed that the oscillation period of the voltage signal equals to 12.5 ms, which is consistent with the simulation (see Appendix J). The decay of the recovered voltage peak is due to loss and disorder effects in the circuit sample. In addition, we can also define the effective occupation density at the *i*th site as $P_i^\alpha = \sum_{l=1}^{L} |V_{i,l}^\alpha|^2$ ($i \neq l, \alpha = 4,5$). Fig. 5(b) illustrates the evolution of the effective occupation density of two fusion degrees, where black and red lines represent results of the fourth and fifth fusion degrees, respectively. It is clearly shown that the mutual transformation of two coupled fusion degrees exists during the BZO, being consistent with the theoretical prediction.

Then, we turn to the fabricated circuit with $F = 2$, and the initial voltage is also given by $V_{3,5}^5(t=0) = 1V$. Fig. 5(c) displays the measured voltage signal of $V_{3,5}^5(t)$. It can be observed that the voltage signal initially decreases, then experiences a slight recovery, reaching its maximum value at *t*=6.284ms. Thus, the period of BO is 6.284ms, which is also consistent with the simulation (see Appendix J). Similarly, the deviation between the experimental and theoretical waveforms is due to the loss and disorder effects in the circuit sample. Additionally, we further present the evolution of the effective occupation density in the circuit simulator, as shown in Fig. 5(d). It is seen that there is a continuous transformation between two fusion degrees, and the experimental measurements are in accord with theoretical results. It is noted that the above experiments only focus on BOs and BZOs of two Fibonacci anyons. Based on the same method, we can also design electric circuits to simulate fusion-dependent BZOs of three Fibonacci anyons.

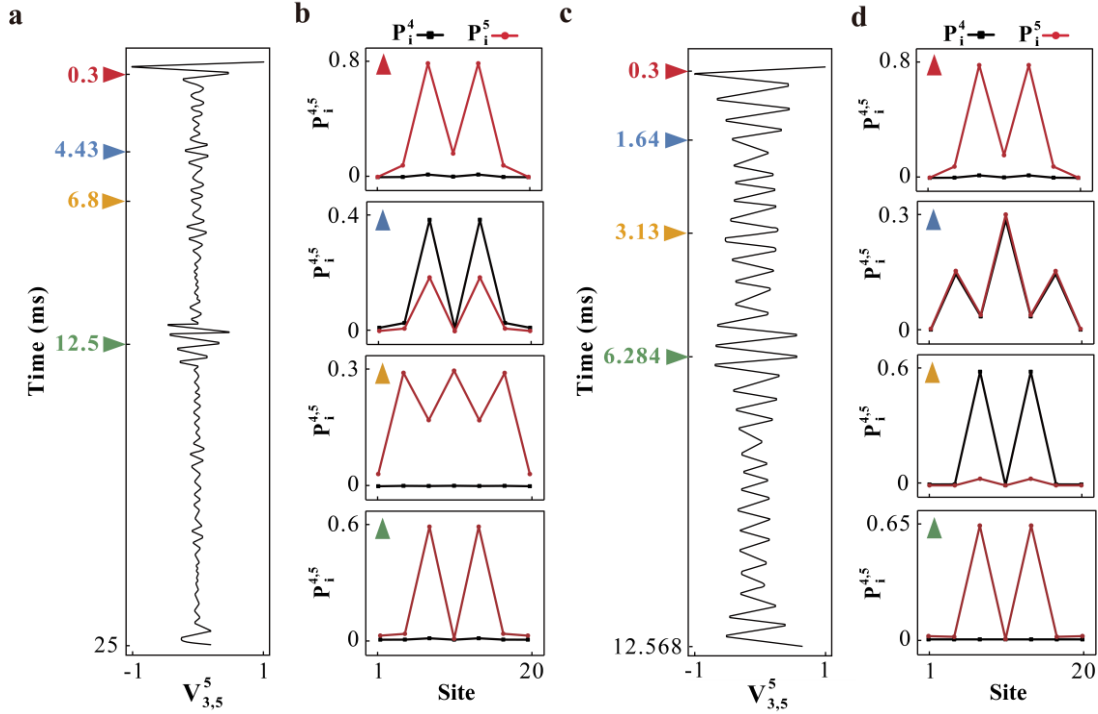

**FIG. 5. Experimental simulation of BZO and BO for two Fibonacci anyons by voltage measurements.** (**a**) The measured voltage signal of $V^5_{3,5}$ in the circuit simulator with $F$=1.5. (**b**) The evolution of the effective occupation density of two fusion degrees in the circuit simulator with $F$=1.5. The black and red lines represent the fourth and fifth fusion degrees, respectively. (**c**) The measured voltage signal of $V^5_{3,5}$ in the circuit simulator with $F$=2. (**d**) The evolution of the effective occupation density of two fusion degrees in the circuit simulator with F=2. Here, the initial voltage is always set as $V^5_{3,5}(t=0) = 1$ V.

## IV. CONCLUSION.

In conclusion, we have proposed a new approach to characterize the non-Abelian fusion of Fibonacci anyons through the utilization of BOs and BZOs. Specifically, we found that by manipulating external forces, it becomes feasible to achieve either a dual or singular Wannier-Stack ladder of two Fibonacci anyons. Remarkably, when a Wannier-Stack ladder emerges alongside BOs, the corresponding values of external force can precisely determine the golden ratio for the fusion matrix. In addition, in scenarios involving three Fibonacci anyons, a pair of internal energy levels arises under a suitable value of external force leading to the manifestation of BZOs of three Fibonacci anyons. Furthermore, we have experimentally fabricated RC circuits to simulate the fusion-dependent BOs and BZOs of two Fibonacci anyons. *It is worth noting that, unlike electrons, Fibonacci anyons do not exhibit a constant force in the presence of an electric field. It is possible to realize an effective external force by linearly increasing the onsite potential along the lattice model of anyons. The potential methods include the strain engineering techniques [50-52] and designing the frequency difference between two coupling optical fields in the*

*cold-atom system [53, 54], which sustain non-Abelian anyons [2].* Our findings establish a profound connection between BOs and the intricate realm of non-Abelian fusion, and provide a versatile platform for simulating a myriad of captivating phenomena associated with the fascinating field of non-Abelian physics.

## ACKNOWLEDGEMENTS

This work is supported by the National Key R & D Program of China under Grant No. 2022YFA1404900, Young Elite Scientists Sponsorship Program by CAST No. 2023QNRC001, and the National Natural Science Foundation of China No. 12104041.

## APPENDIX A: DETAILS ON BASIS VECTORS OF THE SYSTEM WITH TWO FIBONACCI ANYONS.

In this section, we provide a detailed explanation of the basis vectors in our model consisting of two Fibonacci anyons. Fig. 6(a1) illustrates the fusion tree for our 1D lattice model (L=5) with two Fibonacci anyons, denoted as $x_i \in \{I, \tau\}$. Due to the non-Abelian nature of their fusion, there exist multiple possible fusion trees when the positions of these anyons are fixed. As depicted in Fig. 6(b), we present five potential fusion trees corresponding to two Fibonacci anyons located at the first and second sites, resulting in five internal basis vectors of $|x_1, x_2, x_3, x_4, x_5, x_6\rangle = |I, \tau, I, I, I, I\rangle$, $|I, \tau, \tau, I, I, I\rangle$, $|\tau, \tau, I, I, I, I\rangle$, $|\tau, I, \tau, I, I, I\rangle$, $|\tau, \tau, \tau, \tau, \tau\rangle$ induced by non-Abelian fusion.

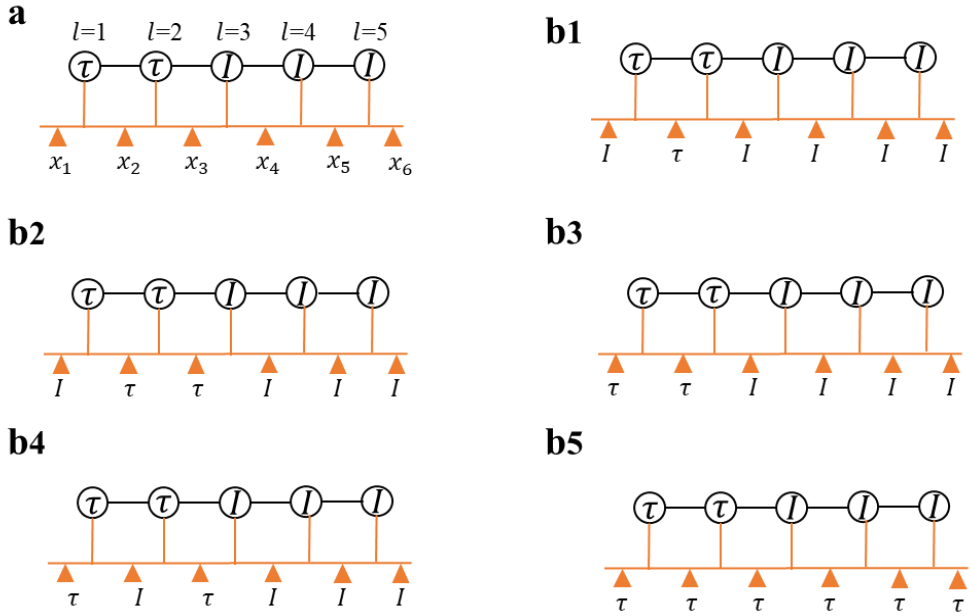

**FIG. 6.** (a). Fibonacci anyonic fusion trees with *L*=5. (b1)-(b5). Five potential fusion trees corresponding to two

Fibonacci anyons located at the first and second sites.

## APPENDIX B: DETAILS ON THE DERIVATION OF THE FUSION MATRIX FOR TWO FIBONACCI ANYONS.

In this section, we give a detailed derivation of the fusion matrix $H_J$. First, let us introduce what the $\mathcal{F}$-matrix is. The $\mathcal{F}$-matrix is a matrix that describes the fusion properties of anyons. In a system composed of multiple anyons, the Hilbert space resulting from the fusion of anyons is independent of the order of fusion. However, different fusion orders can lead to distinct basis vectors for the system. The $\mathcal{F}$-matrix is a unitary transformation matrix that describes the relationship between basis vectors resulting from different fusion orders. In the following, we provide details on deriving the $\mathcal{F}$-matrix.

In the arrangement of three particles from left to right: the leftmost two particles fuse first and then fuse with the rightmost particle, as shown in Fig. 7(a), resulting in five different fusion outcomes and five distinct basis vectors, denoted as Basis 1. Similarly, if the rightmost two particles fuse first, as shown in Fig. 7(b), there are also five possible fusion scenarios, resulting in five basis vectors, denoted as Basis 2. The transformation from Basis 1 to Basis 2 can be accomplished using the $\mathcal{F}$-matrix. In this case, the transformation between different basis vectors can be described using the $\mathcal{F}$-matrix, as shown in Fig. 7(c). Here, $[\mathcal{F}_m^{ijk}]_{pq}$ represents the matrix element of the $\mathcal{F}$-matrix, which can transform the fusion tree with the first (i) and second (j) anyons being fusion initially to the fusion tree with the second (j) and third (k) anyons being fusion initially. Here, these two fusion trees can be expressed as $|((i,j)_p,k)_m\rangle$ and $|(i,(j,k)_q)_m\rangle$. So we have:

$$|((i,j)_p,k)_m\rangle = \sum_q [\mathcal{F}_m^{ijk}]_{pq} |(i,(j,k)_q)_m\rangle, \tag{A1}$$

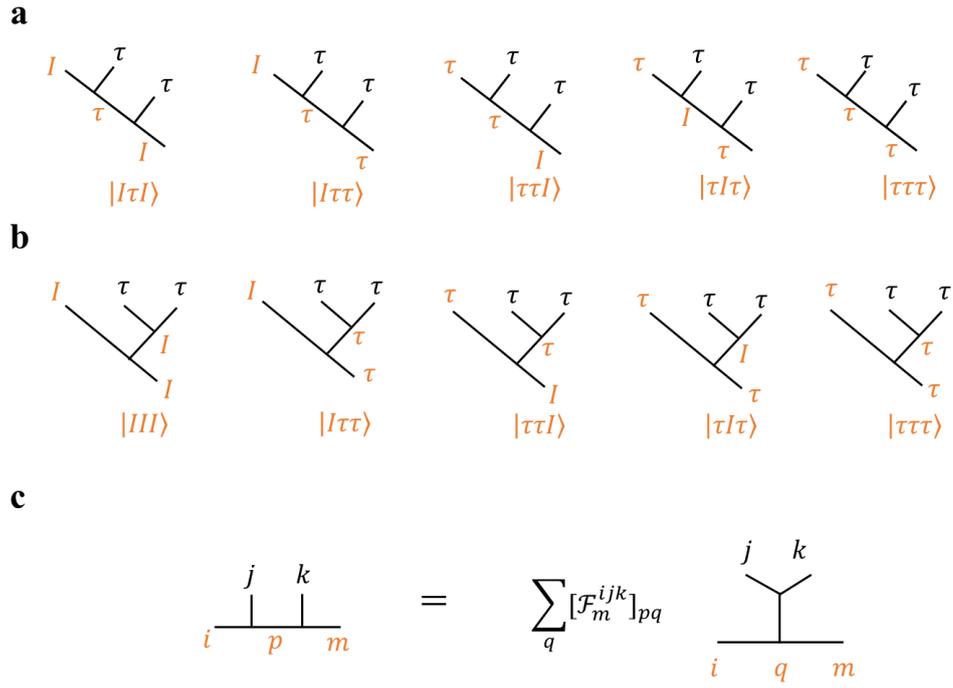

FIG. 7. (a) Schematic diagram of the fusion trees with the leftmost two particles fuse first and then fuse with the rightmost particle. (b) Schematic diagram of the fusion trees with the rightmost two particles fuse first and then fuse with the leftmost particle. (c) The transformation between different basis states.

It has been demonstrated that if at least one of four labels $i$, $j$, $k$ and $m$ of $[\mathcal{F}_m^{ijk}]_{pq}$ is a vacuum state ($I$), as shown in Fig. 8(a), then we always have $[\mathcal{F}_I^{I\tau\tau}]_{\tau I}, [\mathcal{F}_I^{\tau\tau\tau}]_{\tau\tau}$ and $[\mathcal{F}_\tau^{I\tau\tau}]_{\tau\tau} = 1$. In this case, other values of $[\mathcal{F}_m^{ijk}]_{pq}$ are all zero. In addition, when these four labels of $[\mathcal{F}_m^{ijk}]_{pq}$ are all Fibonacci anyons ($\tau$), as shown in Fig. 8(b), the value of $[\mathcal{F}_m^{ijk}]_{pq}$ depends on the detailed fusion role of Fibonacci anyons. Next, we focus on the derivation the $\mathcal{F}$-matrix related to $[\mathcal{F}_\tau^{\tau\tau\tau}]_{pq}$.

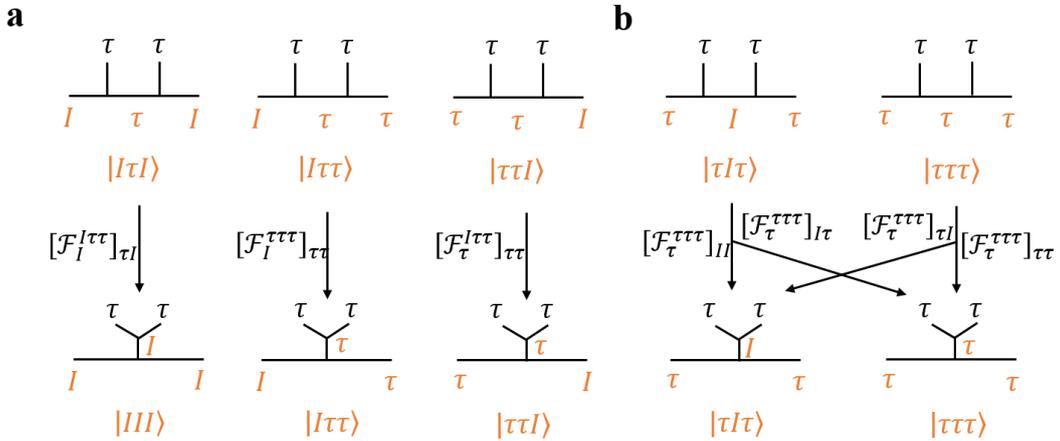

FIG. 8. Schematic representation of the basis transformation resulting from the fusion process of two Fibonacci

anyons.

Firstly, the $\mathcal{F}$-matrix should be a unitary matrix with $\mathcal{F}\mathcal{F}^+ = I$. Thus, we have

$$[\mathcal{F}_\tau^{\tau\tau\tau}]_{II}^2 + [\mathcal{F}_\tau^{\tau\tau\tau}]_{I\tau}^2 = 1, \tag{A2}$$

$$[\mathcal{F}_\tau^{\tau\tau\tau}]_{II}[\mathcal{F}_\tau^{\tau\tau\tau}]_{\tau I} + [\mathcal{F}_\tau^{\tau\tau\tau}]_{I\tau}[\mathcal{F}_\tau^{\tau\tau\tau}]_{\tau\tau} = 0, \tag{A3}$$

$$[\mathcal{F}_\tau^{\tau\tau\tau}]_{\tau I}^2 + [\mathcal{F}_\tau^{\tau\tau\tau}]_{\tau\tau}^2 = 1. \tag{A4}$$

Then, based on Eq. (A1), we have

$$|(\tau,(\tau,(\tau,\tau)_I)_\tau)_I\rangle = [\mathcal{F}_\tau^{\tau\tau\tau}]_{II}|(\tau,((\tau,\tau)_I,\tau)_\tau)_I\rangle + [\mathcal{F}_\tau^{\tau\tau\tau}]_{I\tau}|(\tau,((\tau,\tau)_\tau,\tau)_\tau)_I\rangle. \tag{A5}$$

Using the relationship of $|((\tau,\tau)_\tau,\tau)_\tau\rangle = |(\tau,(\tau,\tau)_\tau)_\tau\rangle$, Eqs. (5) can be re-expressed as:

$$|(\tau,(\tau,(\tau,\tau)_I)_\tau)_I\rangle = [\mathcal{F}_\tau^{\tau\tau\tau}]_{II}|(\tau,((\tau,\tau)_I,\tau)_\tau)_I\rangle + [\mathcal{F}_\tau^{\tau\tau\tau}]_{I\tau}|((\tau,(\tau,\tau)_\tau),\tau)_I\rangle. \tag{A6}$$

Applying Eq. (A1) to Eq. (A6), we have

$$|(\tau,(\tau,(\tau,\tau)_I)_\tau)_I\rangle = [\mathcal{F}_\tau^{\tau\tau\tau}]_{II}\sum_j[\mathcal{F}_\tau^{\tau\tau\tau}]_{Ij}|((\tau,\tau)_j,\tau)_\tau,\tau)_I\rangle + [\mathcal{F}_\tau^{\tau\tau\tau}]_{I\tau}\sum_j[\mathcal{F}_\tau^{\tau\tau\tau}]_{Ij}|((\tau,\tau)_j,\tau)_\tau,\tau)_I\rangle$$

$$= \sum_j([\mathcal{F}_\tau^{\tau\tau\tau}]_{II}[\mathcal{F}_\tau^{\tau\tau\tau}]_{Ij} + [\mathcal{F}_\tau^{\tau\tau\tau}]_{I\tau}[\mathcal{F}_\tau^{\tau\tau\tau}]_{\tau j})|((\tau,\tau)_j,\tau)_\tau,\tau)_I\rangle, \tag{A7}$$

with $j$ being either $I$ or $\tau$. In addition, we also have

$$|(\tau,(\tau,(\tau,\tau)_I)_\tau)_I\rangle = |((\tau,\tau)_I,(\tau,\tau)_I)_I\rangle = |(((\tau,\tau)_I,\tau)_\tau,\tau)_I\rangle. \tag{A8}$$

Comparing the Eq. (A7) and Eq. (A8), we can obtain:

$$[\mathcal{F}_\tau^{\tau\tau\tau}]_{II}[\mathcal{F}_\tau^{\tau\tau\tau}]_{II} + [\mathcal{F}_\tau^{\tau\tau\tau}]_{I\tau}[\mathcal{F}_\tau^{\tau\tau\tau}]_{\tau I} = 1, \tag{A9}$$

$$[\mathcal{F}_\tau^{\tau\tau\tau}]_{I\tau}([\mathcal{F}_\tau^{\tau\tau\tau}]_{II} + [\mathcal{F}_\tau^{\tau\tau\tau}]_{\tau\tau}) = 0. \tag{A10}$$

Similarly, we can obtain:

$$|(\tau,(\tau,(\tau,\tau)_\tau)_I)_\tau\rangle = \sum_j([\mathcal{F}_\tau^{\tau\tau\tau}]_{\tau I}[\mathcal{F}_\tau^{\tau\tau\tau}]_{Ij} + [\mathcal{F}_\tau^{\tau\tau\tau}]_{\tau\tau}[\mathcal{F}_\tau^{\tau\tau\tau}]_{\tau j})|((\tau,\tau)_j,\tau)_I,\tau)_\tau\rangle, \tag{A11}$$

$$|(\tau,(\tau,(\tau,\tau)_\tau)_I)_\tau\rangle = |(((\tau,\tau)_\tau,\tau)_I,\tau)_\tau\rangle. \tag{A12}$$

Finally, the pentagon diagram related to five different fusion trees, as shown in Fig. 9, should also be applied to obtain other required equations of matrix elements $\left[\mathcal{F}_m^{ijk}\right]_{pq}$.

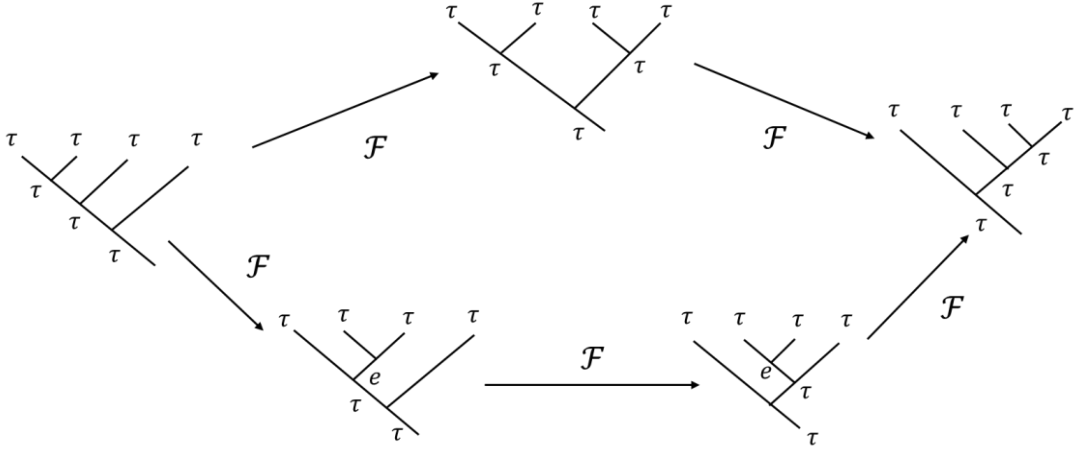

**FIG. 9.** The Pentagon diagram illustration.

The results obtained from the paths in the upper half or the lower half of the pentagon diagram are the same, $\mathcal{F}^2 = \mathcal{F}^3$. From the pentagon equation, we can obtain:

$$[\mathcal{F}_\tau^{\tau\tau\tau}]_{\tau\tau}[\mathcal{F}_\tau^{\tau\tau\tau}]_{\tau\tau} = \sum_e [\mathcal{F}_\tau^{\tau\tau\tau}]_{e\tau}[\mathcal{F}_\tau^{\tau\tau\tau}]_{\tau\tau}[\mathcal{F}_\tau^{\tau\tau\tau}]_{\tau e}$$

$$= [\mathcal{F}_\tau^{\tau\tau\tau}]_{I\tau}[\mathcal{F}_\tau^{\tau I\tau}]_{\tau\tau}[\mathcal{F}_\tau^{\tau\tau\tau}]_{\tau I} + [\mathcal{F}_\tau^{\tau\tau\tau}]_{\tau\tau}[\mathcal{F}_\tau^{\tau\tau\tau}]_{\tau\tau}[\mathcal{F}_\tau^{\tau\tau\tau}]_{\tau\tau}. \tag{A13}$$

This equation is known as the pentagon equation. It's worth noting that, due to the fusion rules, the variable $e$ in the Fig. 9 can take the values of either $I$ or $\tau$. Hence, we have

$$[\mathcal{F}_\tau^{\tau\tau\tau}]_{\tau\tau}{}^2 = [\mathcal{F}_\tau^{\tau\tau\tau}]_{\tau I}[\mathcal{F}_\tau^{\tau\tau\tau}]_{I\tau} + [\mathcal{F}_\tau^{\tau\tau\tau}]_{\tau\tau}{}^3. \tag{A14}$$

Combining Eqs. (A2)-(A14), we can obtain detailed values of $[\mathcal{F}_\tau^{\tau\tau\tau}]_{pq}$ as

$$[\mathcal{F}_\tau^{\tau\tau\tau}]_{\tau\tau} = -\phi^{-1}, \quad [\mathcal{F}_\tau^{\tau\tau\tau}]_{II} = \phi^{-1}, \quad [\mathcal{F}_\tau^{\tau\tau\tau}]_{I\tau} = \phi^{-1/2}, \quad [\mathcal{F}_\tau^{\tau\tau\tau}]_{\tau I} = \phi^{-1/2}, \tag{A15}$$

where $\phi = (\sqrt{5}+1)/2$ is the golden ratio. Thus, the matrix form of the transformation $\mathcal{F}$-matrix between the two sets of basis vectors in Fig. 8 is given by:

$$\mathcal{F} = \begin{pmatrix} 1 & 0 & 0 & 0 & 0 \\ 0 & 1 & 0 & 0 & 0 \\ 0 & 0 & 1 & 0 & 0 \\ 0 & 0 & 0 & \phi^{-1} & \phi^{-1/2} \\ 0 & 0 & 0 & \phi^{-1/2} & -\phi^{-1} \end{pmatrix}. \tag{A16}$$

Using the $\mathcal{F}$-matrix, we can further derive the fusion matrix $H_J$. Here, we consider the case with $L=5$, where two Fibonacci anyons $\tau$ and three vacuum states $I$ exist. Similar to the Heisenberg interaction between spins, the interaction between Fibonacci anyons can be expressed as

$$H_{Heisenberg}^{Fibonacci} = -J \sum_{\langle ij \rangle} \mathcal{P}_{ij}^I, \tag{A17}$$

where $\mathcal{P}_{ij}^I$ is the projection operator acting on the fusion outcome being $I$.

Now, let's write the fusion Hamiltonian for two Fibonacci anyon systems. First, we consider the scenario where two Fibonacci anyons locate at first and second sites. The process of five anyons fusing together is shown in Figs. 10(a)-10(e). For this model, we consider long-range interactions, meaning a tendency for all anyons to fuse into a vacuum state $I$, as shown in Fig. 10(e). To achieve this, we need to perform four basis transformations. The beginning basis is given by $|x_1, x_2, x_3, x_4, x_5, x_6\rangle$ as

$$|I\tau IIII\rangle, \quad |\tau\tau IIII\rangle, \quad |I\tau\tau\tau\tau\tau\rangle, \quad |\tau I\tau III\rangle, \quad |\tau\tau\tau\tau\tau\tau\rangle. \tag{A18}$$

The $\mathcal{F}_1$-matrix is written as

$$\mathcal{F}_1 = \begin{pmatrix} 1 & 0 & 0 & 0 & 0 \\ 0 & 1 & 0 & 0 & 0 \\ 0 & 0 & 1 & 0 & 0 \\ 0 & 0 & 0 & \phi^{-1} & \phi^{-1/2} \\ 0 & 0 & 0 & \phi^{-1/2} & -\phi^{-1} \end{pmatrix}, \tag{A19}$$

changing $|x_1, x_2, x_3, x_4, x_5, x_6\rangle$ to the basis $|x_1, \widetilde{x_2}, x_3, x_4, x_5, x_6\rangle$ of

$$|IIIIII\rangle, \quad |\tau\tau IIII\rangle, \quad |I\tau\tau\tau\tau\tau\rangle, \quad |\tau I\tau\tau\tau\tau\rangle, \quad |\tau\tau\tau\tau\tau\tau\rangle. \tag{A20}$$

Then, $\mathcal{F}_2$-matrix is written as

$$\mathcal{F}_2 = \begin{pmatrix} 1 & 0 & 0 & 0 & 0 \\ 0 & 1 & 0 & 0 & 0 \\ 0 & 0 & 1 & 0 & 0 \\ 0 & 0 & 0 & 1 & 0 \\ 0 & 0 & 0 & 0 & 1 \end{pmatrix}, \tag{A21}$$

changing $|x_1, \widetilde{x_2}, x_3, x_4, x_5, x_6\rangle$ to the basis $|x_1, \widetilde{x_2}, \widetilde{x_3}, x_4, x_5, x_6\rangle$:

$$|IIIIII\rangle, \quad |\tau\tau\tau III\rangle, \quad |I\tau\tau\tau\tau\tau\rangle, \quad |\tau II\tau\tau\tau\rangle, \quad |\tau\tau\tau\tau\tau\tau\rangle. \tag{A22}$$

$\mathcal{F}_3$-matrix is written as

$$\mathcal{F}_3 = \begin{pmatrix} 1 & 0 & 0 & 0 & 0 \\ 0 & 1 & 0 & 0 & 0 \\ 0 & 0 & 1 & 0 & 0 \\ 0 & 0 & 0 & 1 & 0 \\ 0 & 0 & 0 & 0 & 1 \end{pmatrix}, \tag{A23}$$

changing $|x_1, \widetilde{x_2}, \widetilde{x_3}, x_4, x_5, x_6\rangle$ to the basis $|x_1, \widetilde{x_2}, \widetilde{x_3}, \widetilde{x_4}, x_5, x_6\rangle$:

$$|IIIIII\rangle, \quad |\tau\tau\tau\tau II\rangle, \quad |I\tau\tau\tau\tau\tau\rangle, \quad |\tau III\tau\tau\rangle, \quad |\tau\tau\tau\tau\tau\tau\rangle. \tag{A24}$$

$\mathcal{F}_4$-matrix is written as

$$\mathcal{F}_4 = \begin{pmatrix} 1 & 0 & 0 & 0 & 0 \\ 0 & 1 & 0 & 0 & 0 \\ 0 & 0 & 1 & 0 & 0 \\ 0 & 0 & 0 & 1 & 0 \\ 0 & 0 & 0 & 0 & 1 \end{pmatrix}, \tag{A25}$$

changing $|x_1, \widetilde{x_2}, \widetilde{x_3}, \widetilde{x_4}, x_5, x_6\rangle$ to the basis $|x_1, \widetilde{x_2}, \widetilde{x_3}, \widetilde{x_4}, \widetilde{x_5}, x_6\rangle$:

$$|IIIIII\rangle, \quad |\tau\tau\tau\tau\tau I\rangle, \quad |I\tau\tau\tau\tau\tau\rangle, \quad |\tau IIII\tau\rangle, \quad |\tau\tau\tau\tau\tau\tau\rangle. \tag{A26}$$

Combined with the projector $\mathcal{P}^I = diag(1,0,0,1,0)$ the Hamiltonian matrix then becomes

$$H_J = -J\mathcal{F}_1\mathcal{F}_2\mathcal{F}_3\mathcal{F}_4\mathcal{P}^I\mathcal{F}_4\mathcal{F}_3\mathcal{F}_2\mathcal{F}_1 = -J\begin{pmatrix} 1 & 0 & 0 & 0 & 0 \\ 0 & 0 & 0 & 0 & 0 \\ 0 & 0 & 0 & 0 & 0 \\ 0 & 0 & 0 & \phi^{-2} & \phi^{-3/2} \\ 0 & 0 & 0 & \phi^{-3/2} & \phi^{-1} \end{pmatrix}. \quad (A27)$$

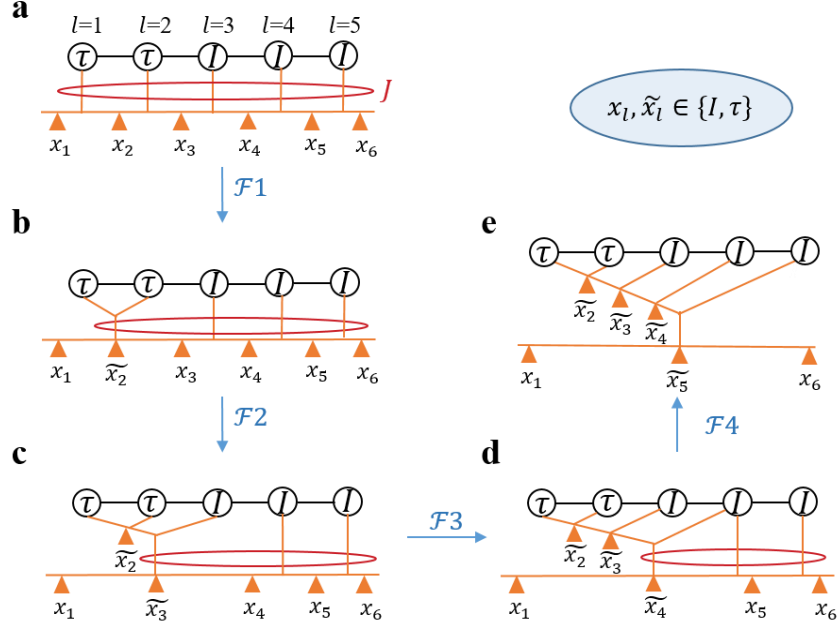

FIG. 10. The fusion trees corresponding to a Fibonacci anyon at the first site and the other Fibonacci anyon at the second site. By applying the $\mathcal{F}_1$-symbol transformation, the fusion chain basis of (a) is transformed into the fusion chain basis of (b). By applying the $\mathcal{F}_2$-symbol transformation, the fusion chain basis of (b) is transformed into the fusion chain basis of (c). By applying the $\mathcal{F}_3$-symbol transformation, the fusion chain basis of (c) is transformed into the fusion chain basis of (d). By applying the $\mathcal{F}_4$-symbol transformation, the fusion chain basis of (d) is transformed into the fusion chain basis of (e). The label $x_l$, $x_{l-1}$ can be either the vacuum particle $I$ or the Fibonacci anyon $\tau$.

Following the same process, we can also consider the scenario where two Fibonacci anyons are at positions one and three. For this model, we also need four basis transformations to achieve the fusion of all anyons into a vacuum state $I$, as shown in Figs. 11(a)-11(e). The beginning basis is given by $|y_1, y_2, y_3, y_4, y_5, y_6\rangle$:

$$|I\tau\tau III\rangle, \quad |\tau\tau\tau III\rangle, \quad |I\tau\tau\tau\tau\tau\rangle, \quad |\tau II\tau\tau\tau\rangle, \quad |\tau\tau\tau\tau\tau\tau\rangle. \quad (A28)$$

The $\mathcal{F}_1$-matrix is written as

$$\mathcal{F}_1 = \begin{pmatrix} 1 & 0 & 0 & 0 & 0 \\ 0 & 1 & 0 & 0 & 0 \\ 0 & 0 & 1 & 0 & 0 \\ 0 & 0 & 0 & 1 & 0 \\ 0 & 0 & 0 & 0 & 1 \end{pmatrix}, \quad (A29)$$

changing the basis $|y_1, y_2, y_3, y_4, y_5, y_6\rangle$ to the basis $|y_1, \widetilde{y_2}, y_3, y_4, y_5, y_6\rangle$ of

$$|I\tau\tau III\rangle, \quad |\tau\tau\tau III\rangle, \quad |I\tau\tau\tau\tau\tau\rangle, \quad |\tau\tau I\tau\tau\tau\rangle, \quad |\tau\tau\tau\tau\tau\tau\rangle. \quad (A30)$$

Then, $\mathcal{F}_2$-matrix is written as

$$\mathcal{F}_2 = \begin{pmatrix} 1 & 0 & 0 & 0 & 0 \\ 0 & 1 & 0 & 0 & 0 \\ 0 & 0 & 1 & 0 & 0 \\ 0 & 0 & 0 & \phi^{-1} & \phi^{-1/2} \\ 0 & 0 & 0 & \phi^{-1/2} & -\phi^{-1} \end{pmatrix}, \tag{A31}$$

changing $|y_1, \widetilde{y_2}, y_3, y_4, y_5, y_6\rangle$ to the basis $|y_1, \widetilde{y_2}, \widetilde{y_3}, y_4, y_5, y_6\rangle$:

$$|I\tau IIII\rangle, \quad |\tau\tau\tau III\rangle, \quad |I\tau\tau\tau\tau\tau\rangle, \quad |\tau\tau I\tau\tau\tau\rangle, \quad |\tau\tau\tau\tau\tau\tau\rangle. \tag{A32}$$

$\mathcal{F}_3$-matrix is written as

$$\mathcal{F}_3 = \begin{pmatrix} 1 & 0 & 0 & 0 & 0 \\ 0 & 1 & 0 & 0 & 0 \\ 0 & 0 & 1 & 0 & 0 \\ 0 & 0 & 0 & 1 & 0 \\ 0 & 0 & 0 & 0 & 1 \end{pmatrix}, \tag{A33}$$

Changing $|y_1, \widetilde{y_2}, \widetilde{y_3}, y_4, y_5, y_6\rangle$ to the basis $|y_1, \widetilde{y_2}, \widetilde{y_3}, \widetilde{y_4}, y_5, y_6\rangle$:

$$|I\tau IIII\rangle, \quad |\tau\tau\tau\tau II\rangle, \quad |I\tau\tau\tau\tau\tau\rangle, \quad |\tau\tau II\tau\tau\rangle, \quad |\tau\tau\tau\tau\tau\tau\rangle. \tag{A34}$$

$\mathcal{F}_4$-matrix is written as

$$\mathcal{F}_4 = \begin{pmatrix} 1 & 0 & 0 & 0 & 0 \\ 0 & 1 & 0 & 0 & 0 \\ 0 & 0 & 1 & 0 & 0 \\ 0 & 0 & 0 & 1 & 0 \\ 0 & 0 & 0 & 0 & 1 \end{pmatrix}, \tag{A35}$$

changing $|y_1, \widetilde{y_2}, \widetilde{y_3}, \widetilde{y_4}, y_5, y_6\rangle$ to the basis $|y_1, \widetilde{y_2}, \widetilde{y_3}, \widetilde{y_4}, \widetilde{y_5}, y_6\rangle$:

$$|I\tau IIII\rangle, \quad |\tau\tau\tau\tau\tau I\rangle, \quad |I\tau\tau\tau\tau\tau\rangle, \quad |\tau\tau III\tau\rangle, \quad |\tau\tau\tau\tau\tau\tau\rangle, \tag{A36}$$

Combined with the projector $\mathcal{P}^I = diag(1,0,0,1,0)$, the Hamiltonian matrix can be expressed as

$$H_J = -J\mathcal{F}_1\mathcal{F}_2\mathcal{F}_3\mathcal{F}_4\mathcal{P}^I\mathcal{F}_4\mathcal{F}_3\mathcal{F}_2\mathcal{F}_1 = -J \begin{pmatrix} 1 & 0 & 0 & 0 & 0 \\ 0 & 0 & 0 & 0 & 0 \\ 0 & 0 & 0 & 0 & 0 \\ 0 & 0 & 0 & \phi^{-2} & \phi^{-3/2} \\ 0 & 0 & 0 & \phi^{-3/2} & \phi^{-1} \end{pmatrix}. \tag{A37}$$

Using the same method, regardless of the positions of the two Fibonacci anyons are, we can derive the fusion Hamiltonian.

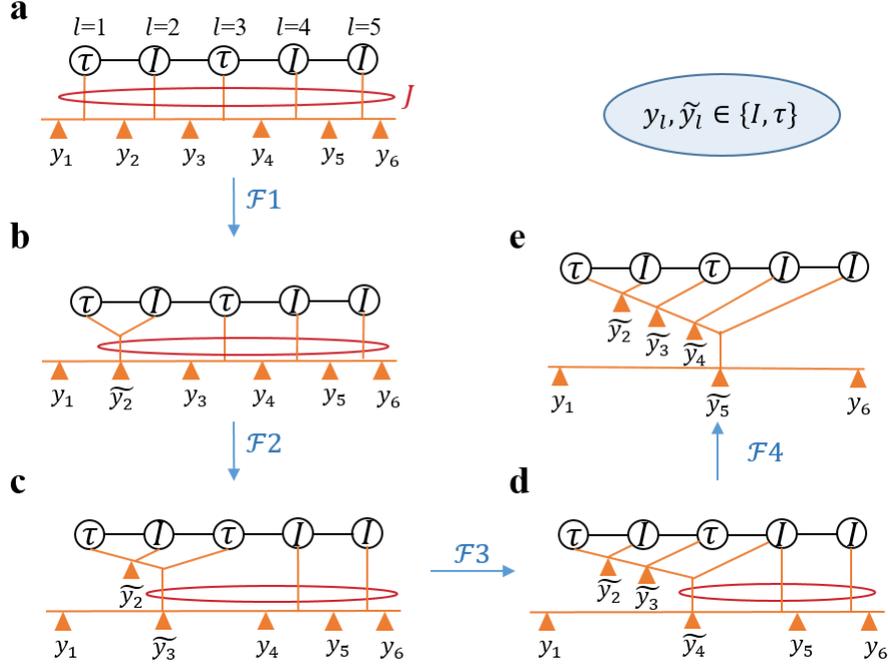

**FIG. 11.** The fusion trees corresponding to a Fibonacci anyon at the first site and the other Fibonacci anyon at the third site. By applying the $\mathcal{F}_1$-symbol transformation, the fusion chain basis of (**a**) is transformed into the fusion chain basis of (**b**). By applying the $\mathcal{F}_2$-symbol transformation, the fusion chain basis of (**b**) is transformed into the fusion chain basis of (**c**). By applying the $\mathcal{F}_3$-symbol transformation, the fusion chain basis of (**c**) is transformed into the fusion chain basis of (**d**). By applying the $\mathcal{F}_4$-symbol transformation, the fusion chain basis of (**d**) is transformed into the fusion chain basis of (**e**). The label $y_l$, $y_{l-1}$ can be either the vacuum particle $I$ or the Fibonacci anyon $\tau$.

## APPENDIX C: THE EVOLUTION OF EIGENENERGIES FOR TWO FIBONACCI ANYONS AS A FUNCTION OF THE EXTERNAL FORCE $F$.

In this section, we calculate the evolution of eigenenergies $\varepsilon_{mn}^1$ related to the fusion degree $C_{m,n}^1$ as a function of F with $J=-1$, $J_t=-1$ and $L=20$, as shown in Fig. 12(a). It is shown that the two-anyon eigenspectra can exhibit the Wannier-Stark ladder $\varepsilon_{mn}^1 = -J + (m+n)F$ at each value of F, and the difference between two contiguous eigen-energies equals to $\delta\varepsilon_1 = F$. Similarly, the eigenspectrum of $C_{m,n}^2$ or $C_{m,n}^3$ is also in the form of a Wannier-Stark ladder with $\varepsilon_{mn}^2 = \varepsilon_{mn}^3 = (m+n)F$. It is noted that the appearance of a Wannier-Stark ladder is expected to induce Bloch oscillation (BO) of two anyons. Next, we calculate the variation of eigen-spectrum $\varepsilon_{mn}^{4,5}$ as a function of the external force, as shown in Fig. 12(b). Clearly, except for the case of F=2 (red nodes) with a single Wannier-Stack appearing, there are always two staggered Wannier-Stack ladders at other values of F, as seen in $F=1.5$ (blue nodes).

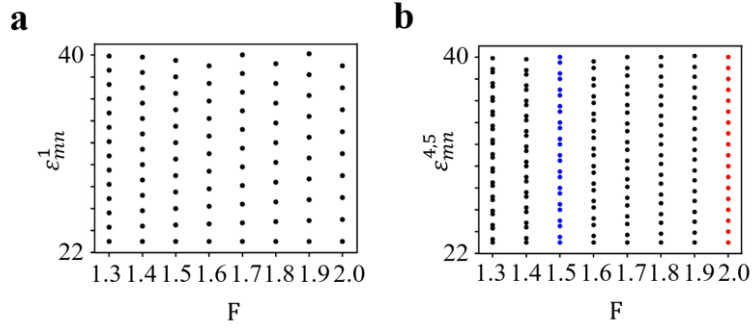

**FIG. 12.** (a) Calculated eigenenergies $\varepsilon_{mn}^1$ related to the fusion degree $C_{m,n}^1$ as a function of $F$. (b) Calculated eigenenergies $\varepsilon_{mn}^{4,5}$ related to the fusion degree $C_{m,n}^{4,5}$ as a function of $F$. The blue and red dots correspond to eigenenergies $\varepsilon_{mn}^{4,5}$ of F=1.5 and F=2, respectively.

## APPENDIX D: DETAILS FOR THE DERIVATION OF THE FUSION MATRIX OF THREE FIBONACCI ANYONS.

In this part, let's derive the fusion Hamiltonian for the three Fibonacci anyon systems. In Supplementary Note 2, we know that the vacuum state $I$ does not affect the fusion Hamiltonian of the Fibonacci anyon system. Therefore, here we only consider the case of three Fibonacci anyons without the vacuum state, as shown in Fig. 13(a). The process of aggregating three Fibonacci anyons together is shown in Fig. 13(a)-13(c). Here, we consider the fusion of three Fibonacci anyons into a vacuum state $I$, as shown in Fig. 13(c). To achieve this, we need perform two basis transformations. The beginning basis is given by $|x_1 x_2 x_3 x_4\rangle$ as

$$|I\tau\tau I\rangle,\ |I\tau I\tau\rangle,\ |I\tau\tau\tau\rangle,\ |\tau I\tau I\rangle,\ |\tau\tau\tau I\rangle,\ |\tau\tau I\tau\rangle,\ |\tau I\tau\tau\rangle,\ |\tau\tau\tau\tau\rangle \qquad (A38)$$

The $\mathcal{F}_1$-matrix is written as

$$\mathcal{F}_1 = \begin{pmatrix} 1 & 0 & 0 & 0 & 0 & 0 & 0 & 0 \\ 0 & 1 & 0 & 0 & 0 & 0 & 0 & 0 \\ 0 & 0 & 1 & 0 & 0 & 0 & 0 & 0 \\ 0 & 0 & 0 & \phi^{-1} & \phi^{-1/2} & 0 & 0 & 0 \\ 0 & 0 & 0 & \phi^{-1/2} & -\phi^{-1} & 0 & 0 & 0 \\ 0 & 0 & 0 & 0 & 0 & 1 & 0 & 0 \\ 0 & 0 & 0 & 0 & 0 & 0 & \phi^{-1} & \phi^{-1/2} \\ 0 & 0 & 0 & 0 & 0 & 0 & \phi^{-1/2} & -\phi^{-1} \end{pmatrix}, \qquad (A39)$$

Changing the basis $|x_1 x_2 x_3 x_4\rangle$ to the basis $|x_1 \widetilde{x_2} x_3 x_4\rangle$ of

$$|I\tau\tau I\rangle,\ |III\tau\rangle,\ |I\tau\tau\tau\rangle,\ |\tau I\tau I\rangle,\ |\tau\tau\tau I\rangle,\ |\tau\tau I\tau\rangle,\ |\tau I\tau\tau\rangle,\ |\tau\tau\tau\tau\rangle. \qquad (A40)$$

Then, $\mathcal{F}_2$-matrix is written as

$$\mathcal{F}_2 = \begin{pmatrix} 1 & 0 & 0 & 0 & 0 & 0 & 0 & 0 \\ 0 & 1 & 0 & 0 & 0 & 0 & 0 & 0 \\ 0 & 0 & 1 & 0 & 0 & 0 & 0 & 0 \\ 0 & 0 & 0 & 1 & 0 & 0 & 0 & 0 \\ 0 & 0 & 0 & 0 & 1 & 0 & 0 & 0 \\ 0 & 0 & 0 & 0 & 0 & \phi^{-1} & 0 & \phi^{-1/2} \\ 0 & 0 & 0 & 0 & 0 & 0 & 1 & 0 \\ 0 & 0 & 0 & 0 & 0 & \phi^{-1/2} & 0 & -\phi^{-1} \end{pmatrix}, \tag{A41}$$

Changing the basis $|x_1 \widetilde{x_2} x_3 x_4\rangle$ to the basis $|x_1 \widetilde{x_2} \widetilde{x_3} x_4\rangle$:

$$|I\tau II\rangle, |II\tau\tau\rangle, |I\tau\tau\tau\rangle, |\tau I\tau I\rangle, |\tau\tau\tau I\rangle, |\tau I\tau\tau\rangle, |\tau\tau I\tau\rangle, |\tau\tau\tau\tau\rangle. \tag{A42}$$

Combined with the projector $\mathcal{P}^{3,I} = diag(1,0,0,0,0,1,0,0)$ the Hamiltonian matrix then becomes

$$H_{3,J} = -J \begin{pmatrix} 1 & 0 & 0 & 0 & 0 & 0 & 0 & 0 \\ 0 & 0 & 0 & 0 & 0 & 0 & 0 & 0 \\ 0 & 0 & 0 & 0 & 0 & 0 & 0 & 0 \\ 0 & 0 & 0 & 0 & 0 & 0 & 0 & 0 \\ 0 & 0 & 0 & 0 & 0 & 0 & 0 & 0 \\ 0 & 0 & 0 & 0 & 0 & \phi^{-2} & \phi^{-2} & -\phi^{-5/2} \\ 0 & 0 & 0 & 0 & 0 & \phi^{-2} & \phi^{-2} & -\phi^{-5/2} \\ 0 & 0 & 0 & 0 & 0 & -\phi^{-5/2} & -\phi^{-5/2} & -\phi^{-3} \end{pmatrix}. \tag{A43}$$

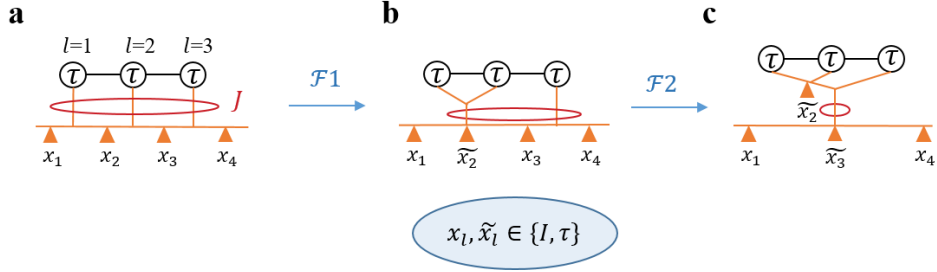

FIG. 13. The process of three Fibonacci anyons fusing together. By applying the $\mathcal{F}_1$-symbol transformation, the fusion chain basis of (**a**) is transformed into the fusion chain basis of (**b**). By applying the $\mathcal{F}_2$-symbol transformation, the fusion chain basis of (**b**) is transformed into the fusion chain basis of (**c**). The label $x_l$, $x_{l-1}$ can be either the vacuum particle $I$ or the Fibonacci anyon $\tau$.

## APPENDIX E: THE TIME-DEPENDENT SCHRÖDINGER EQUATION OF THREE FIBONACCI ANYONS.

In this part, we give the evolution equation for the amplitudes $C_{m,n,q}^{\alpha}$ of three Fibonacci anyons. Here, $C_{m,n,q}^{\alpha}$ is used to represent the probability amplitude of the $\alpha$th fusion degree when the first anyon is at position $m$, the second anyon is at position $n$, and the third anyon is at position $q$. The evolution equation for the amplitude $C_{m,n,q}^{\alpha}$ obtained from the Schrödinger equation $i\partial_t |\psi\rangle = \hat{H}|\psi\rangle$ ($\hbar = 1$) is as followed:

$$i \frac{d}{dt} \boldsymbol{\psi}_{m,n,q} = H_{3,t} [\boldsymbol{\psi}_{m\pm1,n,q} + \boldsymbol{\psi}_{m,n\pm1,q} + \boldsymbol{\psi}_{m,n,q\pm1}] + [H_{3J} + F(m+n+q)I_{8\times8}] \boldsymbol{\psi}_{m,n,q}, \tag{A44}$$

with $\boldsymbol{\psi}_{m,n,q} = [C_{m,n,q}^1, C_{m,n,q}^2, C_{m,n,q}^3, C_{m,n,q}^4, C_{m,n,q}^5, C_{m,n,q}^6, C_{m,n,q}^7, C_{m,n,q}^8]^T$. In the case of the three Fibonacci anyon system, it is observed that there exist eight fusion degrees.

# APPENDIX F: THE EIGENSPECTRA OF THREE FIBONACCI ANYONS AND THE ANALYSIS OF THE OVERALL OSCILLATION PERIOD WITH THREE INTERNAL ENERGY LEVELS.

In this part, our analysis primarily concentrates on the eigenspectra of three Fibonacci anyons and the overall oscillation period with three internal energy levels. The eigenvalues of $H_{3,J}^{6,7,8} = -J \begin{pmatrix} \phi^{-2} & \phi^{-2} & -\phi^{5/2} \\ \phi^{-2} & \phi^{-2} & -\phi^{-5/2} \\ -\phi^{-5/2} & -\phi^{-5/2} & -\phi^{-3} \end{pmatrix}$ equal to $\varepsilon_1 = -0.8937$, $\varepsilon_2 = 0$ and $\varepsilon_3 = 1.4216$, making the internal fusion energy levels become $\delta\varepsilon_f^1 = 0.8937$ and $\delta\varepsilon_f^2 = 1.4216$. In this case, two staggered Wannier-Stack ladders with energy gaps being $\Delta E_1 = F - \delta\varepsilon_f^2 = \delta\varepsilon_f^1$ and $\Delta E_2 = F - \delta\varepsilon_f^1 = \delta\varepsilon_f^2$ can appear with $F=2.3153$. To elucidate this phenomenon further, Fig. 14 illustrates the eigenenergies of three Fibonacci anyons at various values of $F$ with $J=-1$ and $J_t=-1$. It is shown that two staggered Wannier-Stack ladders appear at F=2.3153 (marked by green dots), being consistent with above discussion. For other values of F, the system possesses three internal energy levels, and their oscillation periods should be reconstructed by three periods related to these energy levels. Hence, if the external force satisfies the relationship of $F > 2.3152$, three distinct Wannier-Stack ladders with energy gaps being $\Delta E_1 = \delta\varepsilon_f^1 = 0.8937$, $\Delta E_2 = \delta\varepsilon_f^2 = 1.4216$ and $\Delta E_3 = F - (\delta\varepsilon_f^1 + \delta\varepsilon_f^2)$ are obtained. If $2.3152 > F > 0.8937$, energy gaps of three distinct Wannier-Stack ladders are $\Delta E_1 = F$, $\Delta E_2 = F - \delta\varepsilon_f^1$ and $\Delta E_3 = \delta\varepsilon_f^2$. If $F < 0.8937$, energy gaps of three distinct Wannier-Stack ladders are $\Delta E_1 = \delta\varepsilon_f^1$, $\Delta E_2 = F - \delta\varepsilon_f^1$ and $\Delta E_3 = (\delta\varepsilon_f^1 + \delta\varepsilon_f^2) - F$. The three sub-periods of the system are $T_1 = \frac{2\pi}{\Delta E_1}$, $T_2 = q\frac{2\pi}{\Delta E_3}$ and $T_3 = \frac{2\pi}{\Delta E_3}$. Therefore, the oscillation period of the three Fibonacci anyon system should be the least common multiple of $T_1$, $T_2$, and $T_3$.

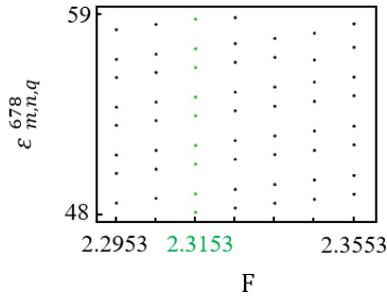

**FIG. 14.** Calculated eigenenergies $\varepsilon_{m,n,q}^{678}$ related to the fusion degree $C_{m,n,q}^{678}$ as a function of F.

# APPENDIX G: DETAILS OF THE DESIGNED *RC* CIRCUIT AND THE DERIVATION OF THE CIRCUIT EIGEN-EQUATIONS.

In this section, we present details of the designed RC circuit and the derivation of the circuit eigen-equations. Circuit nodes $V_{m,n}^4$ and $V_{m,n}^5$ are grounded using a capacitor C and an INIC with effective resistance $R_{m,n}$, while circuit nodes $V_{m,n}^{\prime 4}$ and $V_{m,n}^{\prime 5}$ are grounded using a capacitor C and a regular resistor $R'_{m,n}$. They are respectively marked with green and pink dashed boxes in Fig. 15(a). By appropriately setting the grounding INICs (resistances) as $\frac{1}{R_{m,n}^4} = 4 + \frac{1}{R_{J_1}} + \frac{1}{R_F(m+n)}$ ($\frac{1}{R_{m,n}^5} = 4 + \frac{1}{R_{J_3}} + \frac{1}{R_F(m+n)}$) and grounding resistances as $\frac{1}{R_{m,n}^{\prime 4}} = 4 + \frac{1}{R_{J_1}} + \frac{1}{R_F(m+n)}$ ($\frac{1}{R_{m,n}^{\prime 5}} = 4 + \frac{1}{R_{J_3}} + \frac{1}{R_F(m+n)}$), this satisfies the condition that the diagonal elements of the circuit Hamiltonian $\Pi = i \begin{pmatrix} 0 & -H \\ H & 0 \end{pmatrix}$ are zero. The designed INID for achieving the grounding of node $V_{m,n}^\alpha$ ($R_{m,n}$), single-particle hopping of two Fibonacci anyons ($R_t$), coupled interactions between the fourth and fifth fusion degrees ($R_{J_2}$), as well as self-coupling and external forcing ($R_{J_1}(R_{J_3}) + R_F(m+n)$), is depicted in Fig. 15(b) with arrows in green, black, orange, and blue, respectively. In Fig. 15(c) and (d), we provide complete photographs of the experimental circuits with external forces of 1.5 and 2, respectively. The type of op-amp is LT1363, in which the value of $R_0 = 100\Omega$.

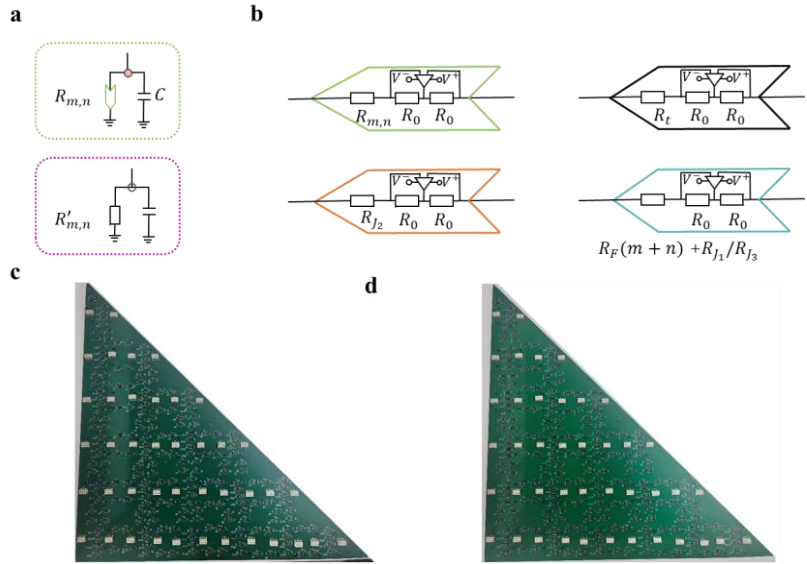

**FIG. 15. (a)** Schematic diagram of the grounding connection in the designed RC circuit. In the pink box are the grounding connections for circuit nodes $V_{m,n}^4$ and $V_{m,n}^5$. In the green box are the grounding connections for f circuit nodes $V_{m,n}^{\prime 4}$ and $V_{m,n}^{\prime 5}$. **(b)** Internal connection structure of INICs. The photographic image of the fabricated circuit (*L*=7) with F=1.5 for **(c)**, and F=2 for **(d)**.

Next, we give a detailed derivation of the circuit eigen-equation. Carrying out the Kirchhoff's law on circuit nodes of $V^4_{m,n}$, $V^5_{m,n}$, $V'^4_{m,n}$, and $V'^5_{m,n}$, we get the following equations as:

$$C\frac{dV^4_{m,n}}{dt} = \frac{1}{R_t}\{(V'^4_{m,n-1} - V^4_{m,n}) + (V'^4_{m,n+1} - V^4_{m,n}) + (V'^4_{m-1,n} - V^4_{m,n}) + (V'^4_{m+1,n} - V^4_{m,n})$$

$$+ \frac{V'^5_{m,n} - V^4_{m,n}}{R_{J_2}} + \left(\frac{1}{R_F(m+n)} + \frac{1}{R_{J_1}}\right)(V'^4_{m,n} - V^4_{m,n}) - \frac{V^4_{m,n}}{R^4_{m,n}}\}, \quad (A45)$$

$$C\frac{dV^5_{m,n}}{dt} = \frac{1}{R_t}\{(V'^5_{m,n-1} - V^5_{m,n}) + (V'^5_{m,n+1} - V^5_{m,n}) + (V'^5_{m-1,n} - V^5_{m,n}) + (V'^5_{m+1,n} - V^5_{m,n})$$

$$+ \frac{V'^4_{m,n} - V^5_{m,n}}{R_{J_2}} + \left(\frac{1}{R_F(m+n)} + \frac{1}{R_{J_3}}\right)(V'^5_{m,n} - V^5_{m,n}) - \frac{V^5_{m,n}}{R^5_{m,n}}\}, \quad (A46)$$

$$C\frac{dV'^4_{m,n}}{dt} = -\frac{1}{R_t}\{(V^4_{m,n-1} - V'^4_{m,n}) + (V^4_{m,n+1} - V'^4_{m,n}) + (V^4_{m-1,n} - V'^4_{m,n}) + (V^4_{m+1,n} - V'^4_{m,n})$$

$$+ \frac{V^5_{m,n} - V'^4_{m,n}}{R_{J_2}} + \left(\frac{1}{R_F(m+n)} + \frac{1}{R_{J_1}}\right)(V^4_{m,n} - V'^4_{m,n}) - \frac{V'^4_{m,n}}{R'^4_{m,n}}\}, \quad (A47)$$

$$C\frac{dV'^5_{m,n}}{dt} = -\frac{1}{R_t}\{(V^5_{m,n-1} - V'^5_{m,n}) + (V^5_{m,n+1} - V'^5_{m,n}) + (V^5_{m-1,n} - V'^5_{m,n}) + (V^5_{m+1,n} - V'^5_{m,n})$$

$$+ \frac{V^4_{m,n} - V'^5_{m,n}}{R_{J_2}} + \left(\frac{1}{R_F(m+n)} + \frac{1}{R_{J_3}}\right)(V^5_{m,n} - V'^5_{m,n}) - \frac{V'^5_{m,n}}{R'^5_{m,n}}\}. \quad (A48)$$

Simplifying the above Eqs. (A45)-(A48), we can obtain

$$\frac{dV^4_{m,n}}{dt} = \frac{1}{CR_t}\{[V'^4_{m,n-1} + V'^4_{m,n+1} + V'^4_{m-1,n} + V'^4_{m+1,n}] + \frac{1}{R_{J_2}}V'^5_{m,n}$$

$$+ (\frac{1}{R_F(m+n)} + \frac{1}{R_{J_1}})V'^4_{m,n} - (4 + \frac{1}{R_F(m+n)} + \frac{1}{R_{J_1}} - \frac{1}{R^4_{m,n}})V^4_{m,n}\}, \quad (A49)$$

$$\frac{dV^5_{m,n}}{dt} = \frac{1}{CR_t}\{[V'^5_{m,n-1} + V'^5_{m,n+1} + V'^5_{m-1,n} + V'^5_{m+1,n}] + \frac{1}{R_{J_2}}V'^4_{m,n}$$

$$+ (\frac{1}{R^5_F(m+n)} + \frac{1}{R_{J_3}})V'^5_{m,n} - (4 + \frac{1}{R_F(m+n)} + \frac{1}{R_{J_3}} - \frac{1}{R^5_{m,n}})V^5_{m,n}\}, \quad (A50)$$

$$\frac{dV'^4_{m,n}}{dt} = \frac{1}{CR_t}\{-[V^4_{m,n-1} + V^4_{m,n+1} + V^4_{m-1,n} + V^4_{m+1,n}] - \frac{1}{R_{J_2}}V^5_{m,n}$$

$$-(\frac{1}{R_F(m+n)} + \frac{1}{R_{J_1}})V^4_{m,n} - (-4 - \frac{1}{R_F(m+n)} - \frac{1}{R_{J_1}} + \frac{1}{R'^4_{m,n}})V'^4_{m,n}\}, \quad (A51)$$

$$\frac{dV'^5_{m,n}}{dt} = \frac{1}{CR_t}\{-[V^5_{m,n-1} + V^5_{m,n+1} + V^5_{m-1,n} + V^5_{m+1,n}] - \frac{1}{R_{J_2}}V^4_{m,n}$$

$$-(\frac{1}{R^5_F(m+n)} + \frac{1}{R_{J_3}})V^5_{m,n} - (-4 - \frac{1}{R_F(m+n)} - \frac{1}{R_{J_3}} + \frac{1}{R'^5_{m,n}})V'^5_{m,n}\}. \quad (A52)$$

In this case, the equation of the voltage evolution can be derived as:

$$\frac{d}{dt}\boldsymbol{V}_{m.n} = \frac{1}{CR_t}\{\boldsymbol{H}'_t[\boldsymbol{V}'_{m\pm1.n} + \boldsymbol{V}'_{m,n\pm1}] + \left[\boldsymbol{H}'_J + \frac{1}{R_F(m+n)}\boldsymbol{I}_{2\times2}\right]\boldsymbol{V}'_{m,n}\}, \quad (A53)$$

$$\frac{d}{dt}\boldsymbol{V}'_{m.n} = -\frac{1}{CR_t}\{\boldsymbol{H}'_t[\boldsymbol{V}_{m\pm1.n} + \boldsymbol{V}_{m,n\pm1}] + \left[\boldsymbol{H}'_J + \frac{1}{R_F(m+n)}\boldsymbol{I}_{2\times2}\right]\boldsymbol{V}_{m,n}\}, \quad (A54)$$

with $\boldsymbol{V}_{m.n} = [V^4_{m,n}, V^5_{m,n}]^T$, $\boldsymbol{V}'_{m.n} = [V'^4_{m,n}, V'^5_{m,n}]^T$, $\boldsymbol{H}'_t = \boldsymbol{I}_{2\times2}$ and $\boldsymbol{H}'_J = \begin{pmatrix} \frac{1}{R_{J_1}} & \frac{1}{R_{J_2}} \\ \frac{1}{R_{J_2}} & \frac{1}{R_{J_3}} \end{pmatrix}$. In this case, the circuit eigen-equation can be written in a matrix form of $i\partial_t|V(t)\rangle = \boldsymbol{\Pi}|V(t)\rangle$ with $|V(t)\rangle =$

$[V^4_{1,2}(t), V^5_{1,2}(t), \ldots, V^4_{L-1,L}(t), V^5_{L-1,L}(t), V'^4_{1,2}(t), V'^5_{1,2}(t), \ldots, V'^4_{L-1,L}(t), V'^5_{L-1,L}(t)]^T$ and $\boldsymbol{\Pi} = i \begin{pmatrix} 0 & H \\ -H & 0 \end{pmatrix}$.

Here, $\boldsymbol{H}$ can be mapped to the Hamiltonian matrix of fourth and fifth fusion degrees for two Fibonacci anyons if the circuit elements satisfy the following relationship of $F = \frac{1}{CR_tR_F(m+n)}$, $\phi^{-2} = \frac{1}{CR_tR_{J_1}}$, $\phi^{-3/2} = \frac{1}{CR_tR_{J_2}}$, and $\phi^{-1} = \frac{1}{CR_tR_{J_3}}$. It is worth noting that the voltage equation in the time domain corresponds to the extended time-dependent Schrödinger equation of two Fibonacci anyon systems expressed as

$$i\partial_t \begin{pmatrix} \boldsymbol{C} \\ \boldsymbol{C'} \end{pmatrix} = \boldsymbol{H_e} \begin{pmatrix} \boldsymbol{C} \\ \boldsymbol{C'} \end{pmatrix}, \tag{A55}$$

with $\boldsymbol{C} = [C^4_{1,2}, C^5_{1,2}, \ldots, C^4_{L-1,L}, C^5_{L-1,L}]^T$, $\boldsymbol{C'} = [C'^4_{1,2}, C'^5_{1,2}, \ldots, C'^4_{L-1,L}, C'^5_{L-1,L}]^T$ and $\boldsymbol{H_e} = i \begin{pmatrix} 0 & H \\ -H & 0 \end{pmatrix}$. It is worth to know that as long as we can obtain the evolution of the state under the action of $\boldsymbol{H_e}$, we can derive the evolution process of the target initial state under the action of $\boldsymbol{H}$. This is because we can expand $H$ into the forms of $H_a$ and $H_e$

$$H_a = \begin{pmatrix} H & 0 \\ 0 & -H \end{pmatrix}, \quad H_e = \begin{pmatrix} 0 & iH \\ -iH & 0 \end{pmatrix}. \tag{A56}$$

$H_a$ and $H_e$ have the same eigenvalues,

$$N^{-1}H_eN = M^{-1}H_aM = \begin{bmatrix} \lambda_1 & \cdots & 0 \\ \vdots & \ddots & \vdots \\ 0 & \cdots & \lambda_n \end{bmatrix}, \tag{A57}$$

where $\lambda_1$ to $\lambda_n$ are eigenvalues of $H_a$ and $H_e$. $M = [W_1 \ W_2 \ \ldots \ W_n]$ and $N = [W'_1 \ W'_2 \ \ldots \ W'_n]$ are the matrixes composed of eigenvectors $W_i$ and $W'_i$ of $H_a$ and $H_e$, respectively. Hence, $H_a$ and $H_e$ can be transformed into each other $P^{-1}H_aP = H_e$ with

$$P = MN^{-1}. \tag{A58}$$

The time-dependent Schrodinger's equation of two anyons is expressed as

$$i\partial_t \psi = H\psi. \tag{A59}$$

Let the state $\psi$ be extended to $\varphi = (\psi \ 0)^T$, we can write the time-dependent Schrodinger's equation of extended Hamiltonian $H_a$ as

$$i\partial_t \begin{pmatrix} \psi \\ 0 \end{pmatrix} = H_a\varphi = \begin{pmatrix} H & 0 \\ 0 & -H \end{pmatrix} \begin{pmatrix} \psi \\ 0 \end{pmatrix} = \begin{pmatrix} H\psi \\ 0 \end{pmatrix}. \tag{A60}$$

The state $\psi$ also evolves under the Hamiltonian $H$. Then, let $\phi = P^{-1}\varphi$, and we have

$$i\partial_t P^{-1}\varphi = P^{-1}H_aPP^{-1}\varphi = H_eP^{-1}\varphi. \tag{A61}$$

So, the state $\phi$ evolves under the Schrödinger equation with the Hamiltonian $H_e$:

$$i\partial_t \phi = H_e \phi. \tag{A62}$$

Therefore, the extended time-dependent Schrödinger equation can simulate the behavior of two Fibonacci anyon systems.

In addition, the eigenfrequencies in the circuit correspond strictly to the eigenvalues in the lattice system. By assuming the harmonic oscillation of voltage signals, we have $V_{m,n}(t) = V_{m,n}e^{iwt}$. Submitting it into Eqs. (A53) and (A54), we have

$$iwV_{m.n} = \frac{1}{CR_t}\{H'_t[V'_{m\pm1.n} + V'_{m,n\pm1}] + \left[H'_J + \frac{1}{R_F(m+n)}I_{2\times2}\right]V'_{m,n}\}, \tag{A63}$$

$$iwV'_{m.n} = -\frac{1}{CR_t}\{H'_t[V_{m\pm1.n} + V_{m,n\pm1}] + \left[H'_J + \frac{1}{R_F(m+n)}I_{2\times2}\right]V_{m,n}\}, \tag{A64}$$

which can be expressed as

$$R_t C\omega \begin{pmatrix} V_{m,n} \\ V'_{m,n} \end{pmatrix} = \begin{pmatrix} 0 & -iH \\ iH & 0 \end{pmatrix}\begin{pmatrix} V_{m,n} \\ V'_{m,n} \end{pmatrix}. \tag{A65}$$

Particularly, the steady-state eigenequation of the circuit simulator can also be mapped to the extended steady-state Schrödinger equation of fourth and fifth fusion degrees for two Fibonacci anyons

$$\varepsilon \begin{pmatrix} C \\ C' \end{pmatrix} = H_e \begin{pmatrix} C \\ C' \end{pmatrix}. \tag{A66}$$

In this case, Eq. (A65) has the same form as the eigenequation of Eq. (A66). Hence, the eigen-energy of two Fibonacci anyons are directly mapped to the eigen-frequency of the circuit as $\varepsilon = R_t C\omega = 2\pi f R_t C$.

### APPENDIX H: SAMPLE FABRICATIONS AND CIRCUIT MEASUREMENTS.

The circuit design is conducted using JLCPCB's EDA program software, involving engineering suitable PCB schematics, stack-up layout, internal layer design, and grounding design. The designed PCB consists of eight layers, including two power layers (±15V), two grounding layers, and all internal layers connected through blind holes. To implement the INIC, we utilize the LT1363 operational amplifier (OpAmp), which are powered by external voltages of ±15V. All capacitors and resistors are packaged in 0603 form factor for compactness. Additionally, a WK6500B impedance analyzer is employed to select circuit elements with high accuracy (disorder strength only 1%). The values of all circuit elements are sufficiently large to neglect any influence from effective resistances or parasitic capacitances in the circuit sample. As for the time-domain voltage measurement, one circuit node needs to be excited, and we set the input signal of this circuit node to 1V. In this case, when initializing the circuit in experiments, each node should be connected to an external voltage signal of 1V or 0. We use the relay model G6K

(Omron) to connect the circuit nodes and the external voltage sources. The relays are controlled by a mechanical switch through a 5V signal. With this setting, external signals can be removed simultaneously. Thus, we connect the nodes to an oscilloscope via coaxial cables and measure the voltage signal after turning off the switch. Additionally, a 4-channel oscilloscope DSO7104B (Agilent Technologies) is used in experiments to collect time-domain voltage signals. For each circuit, we conducted four rounds of measurements, with at least twenty measurements per round, to verify the reproducibility of the obtained results.

**APPENDIX I: SIMULATION RESULTS OF IMPEDANCE RESPONSES FOR CIRCUIT SIMULATORS.**

In this section, we give the simulations results of impedance responses for circuit simulators using the LTSpice software. The circuit dimension designed here is L=7, which is equivalent to the dimensions used in Fig. 4(c) in the main text. Firstly, we simulate the circuit simulator with $R_F = 1500\Omega$ (F=1.5). Impedance spectra of three circuit nodes $Z_{3,4}^5$, $Z_{3,5}^4$ and $Z_{3,5}^5$ are illustrated by blue, black and red lines in Fig. 16(a). There are two intervals between adjacent peaks, equaling to $\delta f_1$ =0.16kHz and $\delta f_2$=0.08kHz, respectively. This is consistent with the existence of two Wannier Stark ladders in the mapped lattice model with $\delta\varepsilon_1$=1 and $\delta\varepsilon_2$=0.5. We also simulate the other circuit with $R_F = 2000\Omega$, corresponding to F=2. Fig. 16(b) presents the simulated impedance spectra of $Z_{3,4}^5$, $Z_{3,5}^4$ and $Z_{3,5}^5$. We can observe the interval between two adjacent impedance peaks are equally spaced with $\delta f$=0.16kHz. This phenomenon indicates the existence of a single Wannier Stark ladder in the mapped lattice model, being consistent with theoretical analysis with $\delta\varepsilon$=1. It is important to note that a little deviation of the frequencies of some peaks is due to the finite size effect.

Then, we showcase the simulated impedance spectrum of the circuit simulator with L=20. Figs. 16(c) and 16(d) show summed impedances of a total of eighteen circuit nodes, from $Z_{12,5}^4$ to $Z_{15,7}^5$, with $R_F = 1500\Omega$ and $R_F = 2000\Omega$, respectively It is shown that there is no deviation of frequency peaks resulting from the finite size effect.

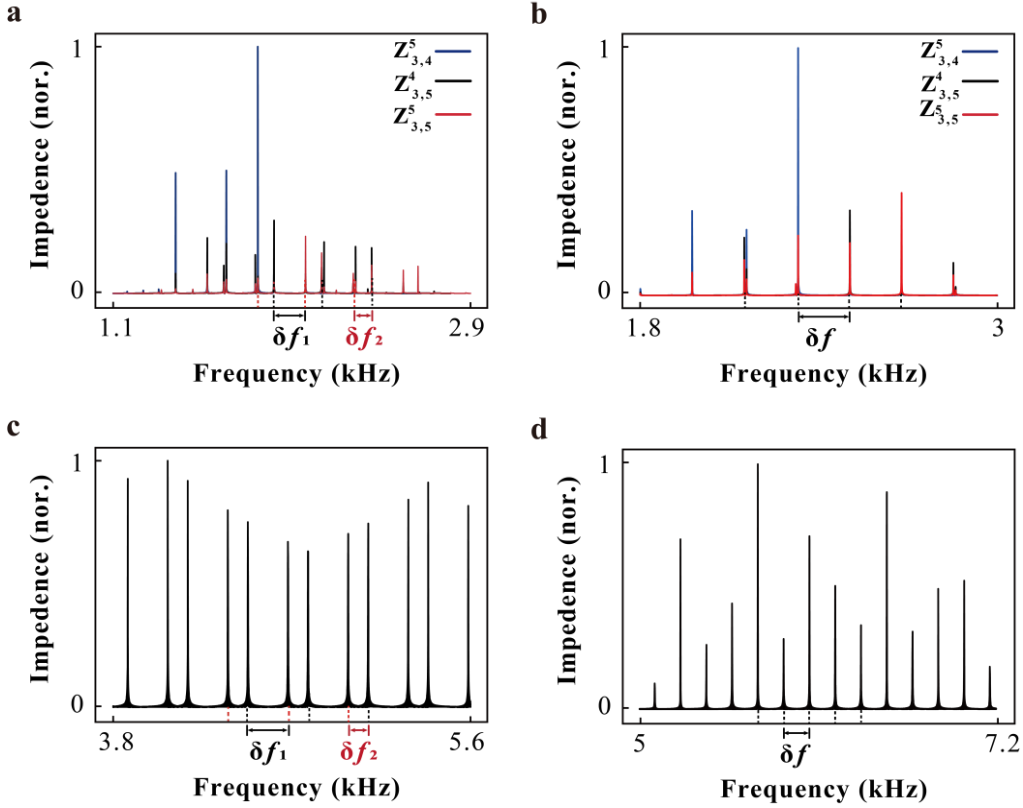

**FIG. 16.** The simulated impedance spectra of $Z^5_{3,4}$, $Z^4_{3,5}$ and $Z^5_{3,5}$ for circuit simulators ($L$=7) with $F$=1.5 for **(a)**, and $F$=2 for **(b)**. The sum of simulated impedance spectra for eighteen circuit nodes (from $Z^4_{12,5}$ to $Z^5_{15,7}$) of circuit simulators ($L$=20) with F=1.5 for **(c)**, and F=2 for **(d)**.

## APPENDIX J: SIMULATED VOLTAGE DYNAMICS OF THE TWO FIBONACCI ANYON CIRCUIT SIMULATOR.

In this section, we will show the simulation of voltage dynamics of the two Fibonacci anyon circuit simulator using LTSpice software. The size if designed circuits is set as L=7, which is equivalent to the dimensions used in Fig. 5(a) in the main text. Firstly, we simulate the voltage evolution in the designed circuit with $F$=1.5. The initial voltage is set as $V^5_{3,5}(t=0) = 1V$. Fig. 17(a) shows the simulated voltage signal of $V^5_{3,5}$. It can be observed that the oscillation period of the voltage signal equals to 12.56ms. Then, we turn to the designed circuit with $F=2$, and the initial voltage is also given by $V^5_{3,5}(t=0) = 1V$. Fig. 17(b) displays the simulated voltage signal of $V^5_{3,5}(t)$. Next, we calculate the time dynamics of the circuit with $F=2$, where the initial voltage distribution is given by $V^5_{3,5}(t=0) = 1V$. The result shows the oscillation period of the voltage signal equals to 6.28ms.

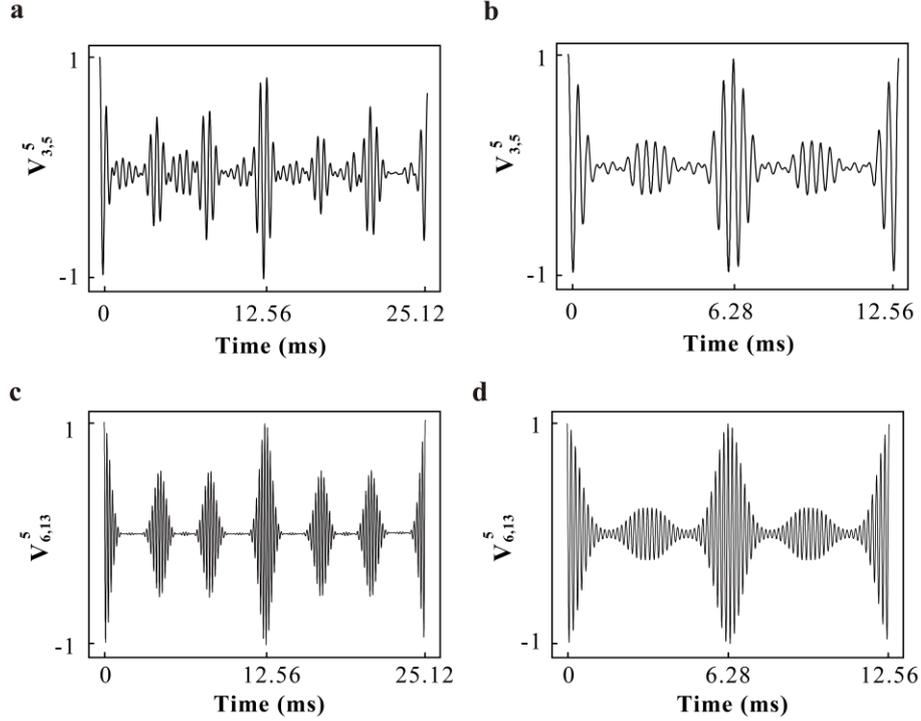

**FIG. 17.** The simulated voltage signals of $V_{3,5}^5$ in the designed RC circuits (L=7) with F=1.5 for **(a)**, and F=2 for **(b)**. The simulated voltage signals of $V_{6,13}^5$ in the designed RC circuits (L=20) with F=1.5 for **(c)**, and F=2 for **(d)**.

To evaluate the influence of the finite size effect, we simulate the voltage dynamics when the size of the circuit simulator is increased. Here, the size of the designed circuits is set as L=20. Firstly, we simulate the voltage evolution in the designed circuit with $F$=1.5. The initial voltage is set as $V_{6,13}^5(t=0) = 1V$. Fig. 17(c) displays the measured voltage signal of $V_{6,13}^5$. We can find the oscillation period of the voltage signal equals to 12.56ms. Next, we turn to the designed circuit with $F=2$, and the initial voltage is also given by $V_{6,13}^5(t=0) = 1V$. Fig. 17(d) shows the simulated voltage signal of $V_{3,5}^5$. It can be observed that the oscillation period of the voltage signal equals to 6.28ms. From about results, we can see that the oscillation periods are nearly unchanged with $L$=7 and 20.